\documentclass[iop]{emulateapj}
\usepackage{url}
\usepackage{stmaryrd}
\usepackage[colorinlistoftodos]{todonotes}
\usepackage{soul}
\usepackage{color}

\definecolor{light-gray}{gray}{0.85}
\usepackage{gensymb}

\slugcomment{}

\shorttitle{The Sizes of z=9-10 Candidate Galaxies}
\shortauthors{Holwerda et al.}

\begin{document}

\title{
The Sizes of Candidate $z\sim9-10$ Galaxies:\\
confirmation of the bright CANDELS sample and relation with luminosity and mass.
}

\author{
B.W. Holwerda\altaffilmark{1*}, 
R. Bouwens\altaffilmark{1},
P. Oesch\altaffilmark{2}, 
R. Smit\altaffilmark{1}, 
G. Illingworth\altaffilmark{3},
I. Labbe\altaffilmark{1}
% M. Franx -  pay for it...
}

\altaffiltext{*}{E-mail: holwerda@strw.leidenuniv.nl, twitter: @benneholwerda}
\altaffiltext{1}{Leiden Observatory, Leiden University, P.O. Box 9513, 2300 RA Leiden, The Netherlands}
\altaffiltext{2}{Yale Center for Astronomy and Astrophysics, Yale University, New Haven, CT 06520}
\altaffiltext{3}{UCO/Lick Observatory, University of California, Santa, Cruz, CA 95064, USA}

\begin{abstract}

Recently, a small sample of six $z\sim9-10$ candidates was discovered in CANDELS that are $\sim10-20\times$ more luminous than any of the previous $z\sim9-10$ galaxies identified over the HUDF/XDF and CLASH fields. We measure the sizes of these candidates to map out the size evolution of galaxies from the earliest observable times. Their sizes are also used to provide a valuable constraint on whether these unusual galaxy candidates are at high redshift. Using galfit to derive sizes from the CANDELS F160W images of these candidates, we find a mean size of $0\farcs13\pm0\farcs02$ (or 0.5$\pm$0.1 kpc at $z\sim9-10$).
This handsomely matches the 0.6 kpc size expected extrapolating lower redshift measurements to $z\sim9-10$, while being much smaller than the 0\farcs59 mean size for lower-redshift interlopers to $z\sim9-10$ photometric selections lacking the blue IRAC color criterion. This suggests that source size may be an effective constraint on contaminants from $z\sim9-10$ selections lacking IRAC data. Assuming on the basis of the strong photometric evidence that the Oesch et al. 2014 sample is entirely at $z\sim9-10$, we can use this sample to extend current constraints on the size-luminosity, size-mass relation, and size evolution of galaxies to $z\sim10$. We find that the $z\sim9-10$ candidate galaxies have broadly similar sizes and luminosities as $z\sim6$-8 counterparts with star-formation-rate surface densities in the range of $\rm \Sigma_{SFR}=1-20\, M_\odot~ yr^{-1}\, kpc^{-2}$.  The stellar mass-size relation is uncertain, but shallower than those inferred for lower-redshift galaxies. In combination with previous size measurements at z=4-7, we find a size evolution of $(1+z)^{-m}$ with $m=1.0\pm0.1$ for $>0.3L^*_{z=3}$ galaxies, consistent with the evolution previously derived from $2 < z < 8$ galaxies.
\end{abstract}


\section{\label{s:intro}Introduction}

The installation of the WFC3/IR camera on the {\em Hubble Space Telescope (HST)} has revolutionized the search for high-redshift ($z>6$) galaxies. At present, some $\gtrsim800 ~ z=6-8$ galaxies are now known \citep{Bouwens14}, from deep, wide-area searches over the Hubble UltraDeep Field \citep[HUDF,][]{hudf}, the WFC3 Early Release Survey \citep[ERS,][]{Windhorst11}, the CANDELS project \citep{Grogin11, Koekemoer11}, and the Brightest of Reionizing Galaxies \citep[BoRG,][]{Trenti11,Trenti12,Trenti12a,Bradley12} fields. 

The high-redshift frontier has now moved to $z\sim9-10$, with a dozen high-fidelity candidates known \citep{Zheng12, Coe13, Bouwens11b, Bouwens11a, Bouwens13, Ellis13, Oesch13,Oesch14}. These highest redshift candidates can be identified by their extremely red near-infrared colors ($J-H >$ 0.5), a lack of flux in bluer bands, and blue $H-4.5\mu$m colors. The first $z\sim9-10$ candidates were found both behind lensing clusters \citep[e.g.,][]{Coe13, Zheng12}, and in ultra-deep WFC3/IR observations \citep{Bouwens11b,Ellis13, Oesch13}.

While most of the initial $z\sim9-10$ candidates were intrinsically very faint, \cite{Oesch14} recently discovered a small sample of bright galaxy candidates over the CANDELS North and South. Remarkably, the \cite{Oesch14} candidates had luminosities that were some 10-20$\times$ brighter than the candidates discovered over the HUDF/XDF or behind lensing clusters, potentially raising questions about their high-redshift nature and whether the candidates are actually at $z\sim9-10$.

One way of testing the high-redshift nature of these candidates is by measuring their sizes and comparing these sizes against expectations for luminous galaxy candidates at $z\sim9-10$, as well as the sizes of potential interlopers to $z\sim9-10$ selections.  The analytical models from \cite{Fall80} and \cite{Mo98} predict effective radii should scale with redshift somewhere between $(1+z)^{-1}$ for galaxies living in halos of fixed mass or $(1+z)^{-1.5}$ at a fixed circular velocity. Observational evidence from earlier samples also points to such scaling relations, with some studies preferring $(1+z)^{-1}$ \citep{Bouwens04,Bouwens06,Oesch10b}, some studies preferring $(1+z)^{-1.5}$ \citep{Ferguson04}, and some studies lying somewhere in between \citep{Hathi08, Ono13, Shibuya15}.\footnote{After submission of this manuscript another analysis appeared on the arXiv \citet{Curtis-Lake14}, which claims to find no evidence for evolution in galaxy sizes across z=4-8. This seems to be in tension with all of the previous literature and is also inconsistent with the recent studies by \citet{Kawamata14} and \citet{Shibuya15} and what is found from stacking ultra-deep observations of galaxies \citep[see Figure 22 in][]{Bouwens14}.}. {For a comprehensive overview of the size relations of galaxies observed with HST, we refer the reader to \protect\cite{Shibuya15}. }

While clearly the strongest evidence for the bright candidates from \cite{Oesch14} being at $z\sim9-10$ would seem to be from the photometric constraints,
a measurement of their sizes can serve as a useful sanity check on their high-redshift nature.  Such tests are useful given the history of the former $z\sim10$ candidate UDFj-39546284 identified in the HUDF09 observations \citep{Bouwens11}, but which subsequent data suggests is more likely an extreme emission-line galaxy at $z\sim2$ based on the non-detection of the candidate in the $JH_{140}$ observations \citep{Ellis13, Bouwens13} and the tentative detection of an emission line at $\sim$1.6$\mu$m \citep{Brammer13}.

\begin{figure}
\begin{center}
\includegraphics[width=0.5\textwidth]{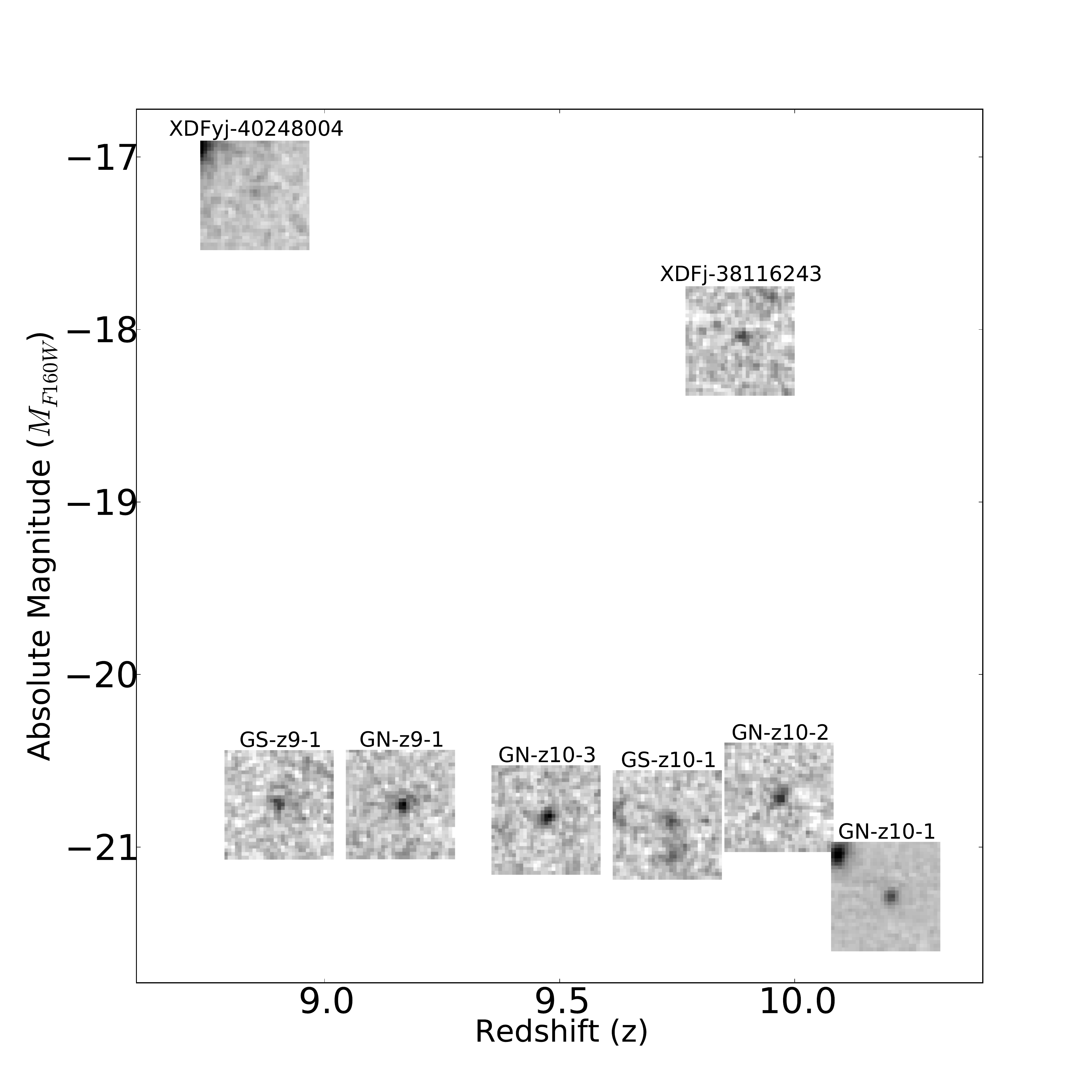}
\caption{{\em F160W} cutouts of our eight $z\sim9-10$ candidate galaxies presented as a function of their respective redshifts and absolute magnitudes to highlight the separation in luminosity between current CANDELS and XDF samples.}
\label{f:overview}
\end{center}
\end{figure}

Additionally, assuming on the basis of the strong photometric evidence  that the \cite{Oesch14} candidates are indeed all bona-fide $z\sim9-10$ galaxies (Figure \ref{f:overview}), the luminosity and redshift of the sources provide the opportunity to constrain the size evolution of luminous galaxies to $z\sim10$, for the first time, and also pursue an exploration of the relation between size and luminosity or mass relation at $z\sim9-10$.  Previously studied $z\sim9-10$ samples \citep{Ono13} consisted almost entirely of extremely faint sources with smaller, more uncertain sizes, making it difficult to optimally constrain the size evolution to $z=9-10$ galaxies.

The purpose of this paper is to
(1) measure the sizes of the candidate $z=9$-10 galaxies reported in \cite{Oesch14} to test if these sources are consistent with corresponding to star-forming galaxies at $z\sim9-10$,
(2) characterize the size evolution of luminous galaxies to the highest-accessible redshifts, and 
(3) explore any change in their star-formation-rate (SFR) surface density of galaxies from the earliest accessible epoch.  
A measurement of the size distribution of $z\sim10$ galaxies is critical for design of current and future observations with the James Webb Space Telescope (JWST), Atacama Large Millimeter Array (ALMA), and the Extremely-Large Telescopes (ELTs).
We adopt $\Omega_M=0.3, \Omega_\Lambda=0.7, H_0=70\, km\, s^{-1} \, Mpc^{-1}$, consistent with recent WMAP9 \citep[][]{Hinshaw13} or Planck results \citep{planck}. We express galaxy UV luminosities in units of the characteristic luminosity $(L^*_{z=3})$ at $z\sim3$, i.e., $M_{1600}(z=3)=-21.07$ \citep{Steidel99}.

\begin{figure*}
\begin{center}
\begin{minipage}{0.11\linewidth}
\begin{center}
% GNDJ-6254514316
GN-z10-1	\\
$m_{H} = 25.95\pm0.07$\\
$r_e=0.6$ kpc 
\end{center}
\end{minipage}
\begin{minipage}{0.11\linewidth}
\begin{center}
% GSDJ-2320550417
GS-z9-1\\
$m_{H}=26.60\pm0.20$\\
$r_e=0.8$ kpc
\end{center}
\end{minipage}
\begin{minipage}{0.11\linewidth}
\begin{center}
% GNDJ-6522418427
GN-z9-1
$m_{H}=26.62\pm0.14$\\
$r_e=0.6$ kpc
\end{center}
\end{minipage}
\begin{minipage}{0.11\linewidth}
\begin{center}
% GNWJ-6040814299
GN-z10-3	
$m_{H}=26.76\pm0.15$\\
$r_e=0.4$ kpc
\end{center}
\end{minipage}
\begin{minipage}{0.11\linewidth}
\begin{center}
% GSDJ-2269746283
GS-z10-1\\
$m_{H}=26.90\pm0.20$\\
$r_e=0.5$ kpc
\end{center}
\end{minipage}
\begin{minipage}{0.11\linewidth}
\begin{center}
% GNDJ-7227314227
GN-z10-2\\
$m_{H}=26.81\pm0.14$\\
$r_e=0.5$ kpc
\end{center}
\end{minipage}
\begin{minipage}{0.11\linewidth}
\begin{center}
XDFj-38116243\\
$m_{H}=29.87\pm0.40$\\
$r_e=0.3$ kpc
\end{center}
\end{minipage}
\begin{minipage}{0.11\linewidth}
\begin{center}
XDFyj-40248004\\
$m_{H}=29.87\pm0.30$\\
$r_e=0.5$ kpc
\end{center}
\end{minipage}\hfill
\begin{minipage}{0.11\linewidth}
\begin{center}
% GNDJ-6254514316
% GN-z10-1	\\
\includegraphics[width=\textwidth]{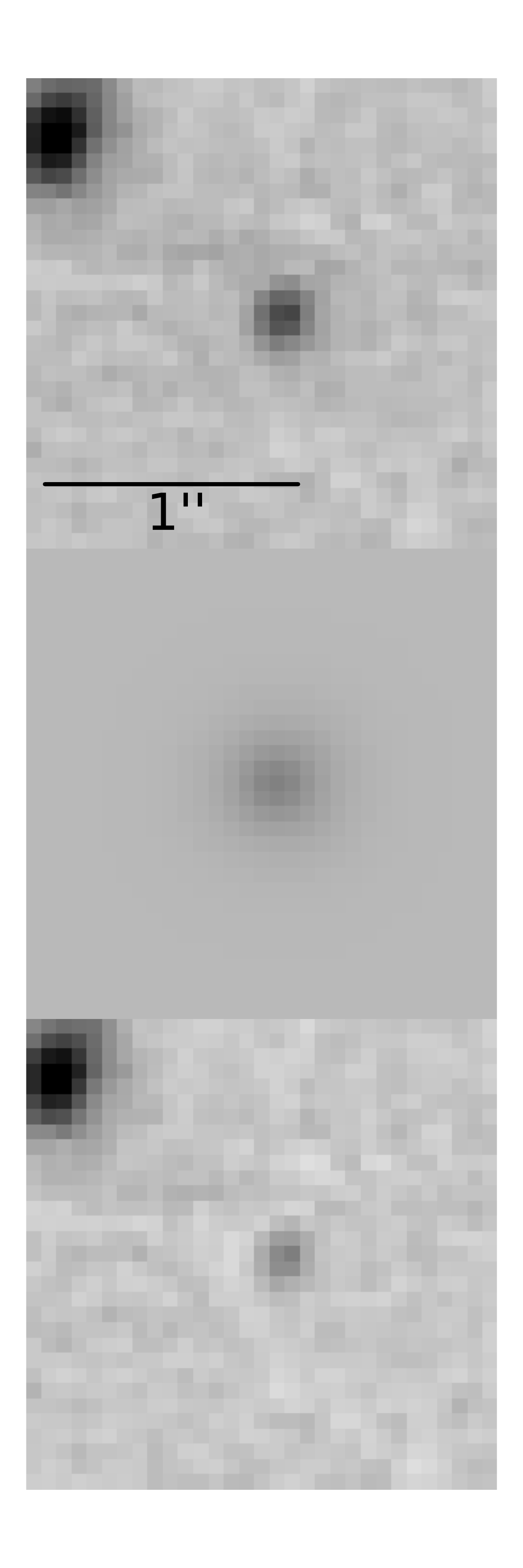}
\end{center}
\end{minipage}
\begin{minipage}{0.11\linewidth}
\begin{center}
% GSDJ-2320550417
% GS-z9-1\\
\includegraphics[width=\textwidth]{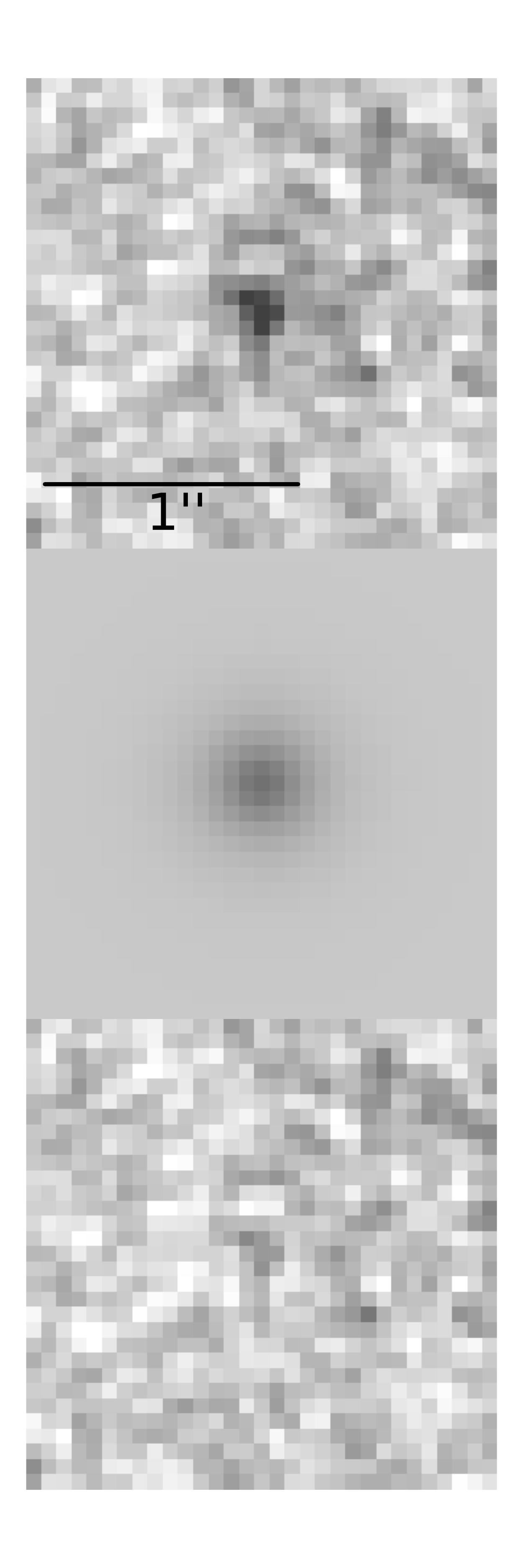}
\end{center}
\end{minipage}
\begin{minipage}{0.11\linewidth}
\begin{center}
% GNDJ-6522418427
% GN-z9-1
\includegraphics[width=\textwidth]{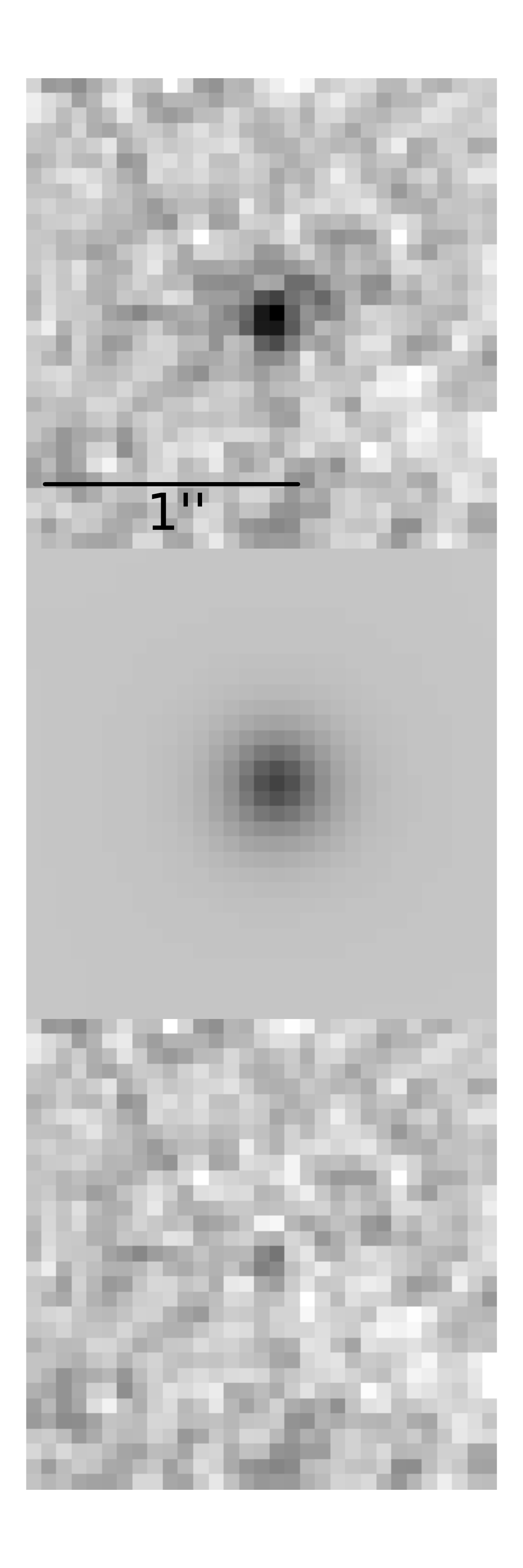}
\end{center}
\end{minipage}
\begin{minipage}{0.11\linewidth}
\begin{center}
% GNWJ-6040814299
% GN-z10-3	
\includegraphics[width=\textwidth]{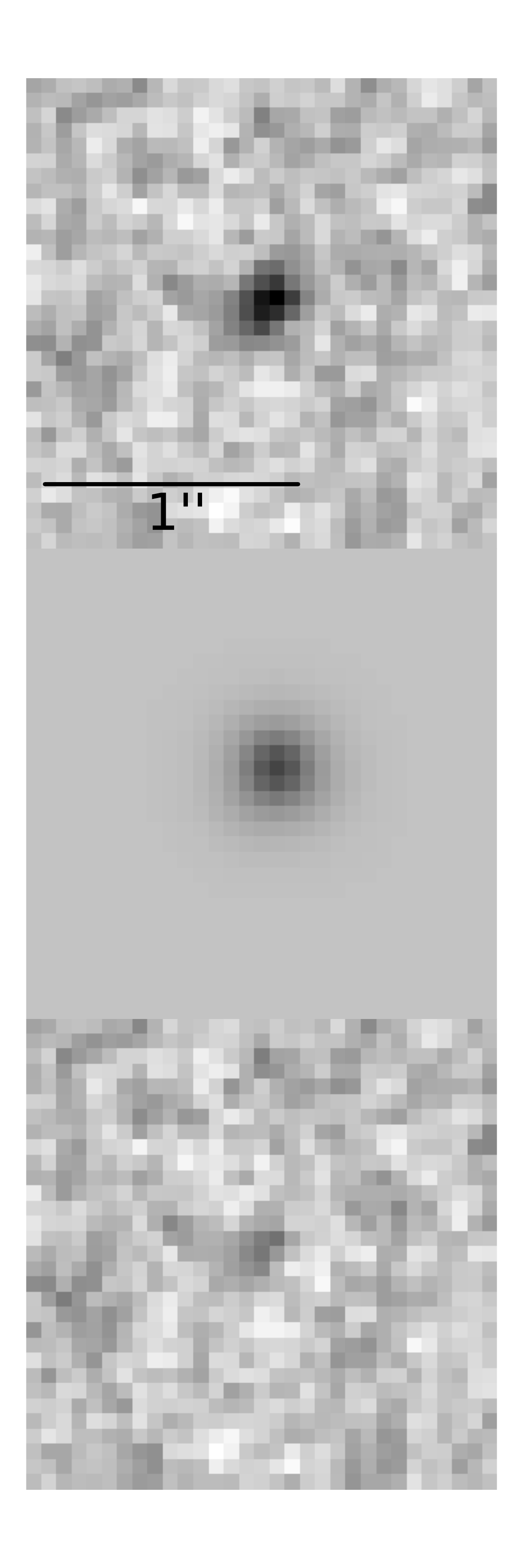}
\end{center}
\end{minipage}
\begin{minipage}{0.11\linewidth}
\begin{center}
% GSDJ-2269746283
% GS-z10-1\\
\includegraphics[width=\textwidth]{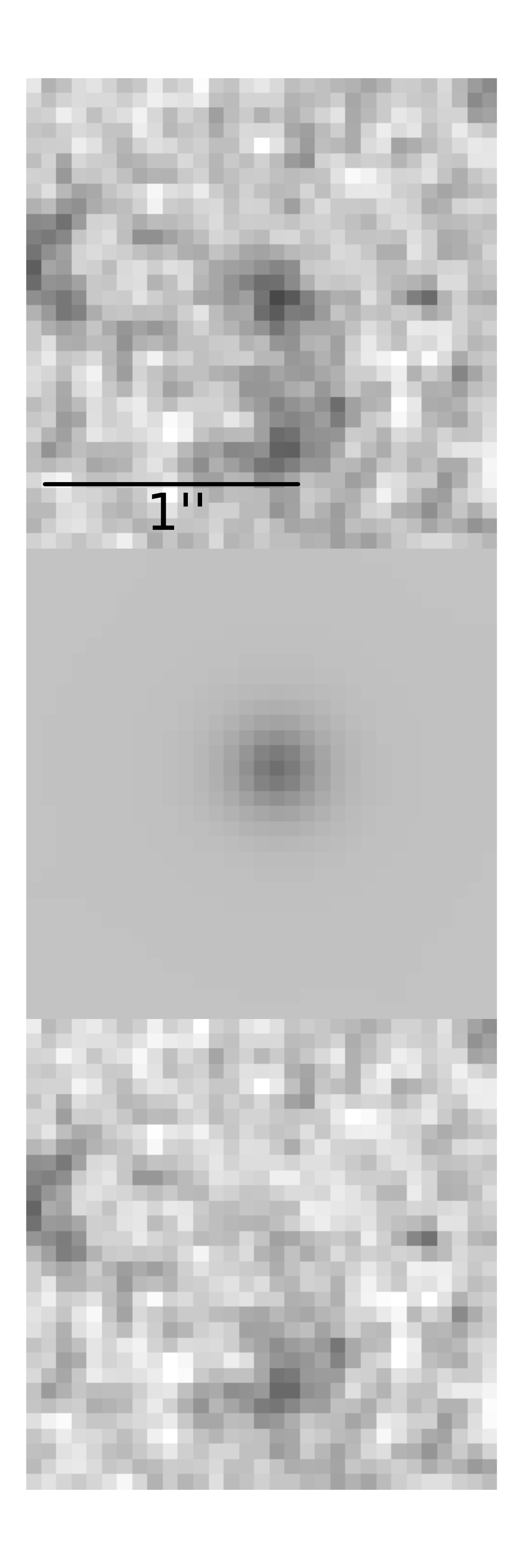}
\end{center}
\end{minipage}
\begin{minipage}{0.11\linewidth}
\begin{center}
% GNDJ-7227314227
% GN-z10-2\\
\includegraphics[width=\textwidth]{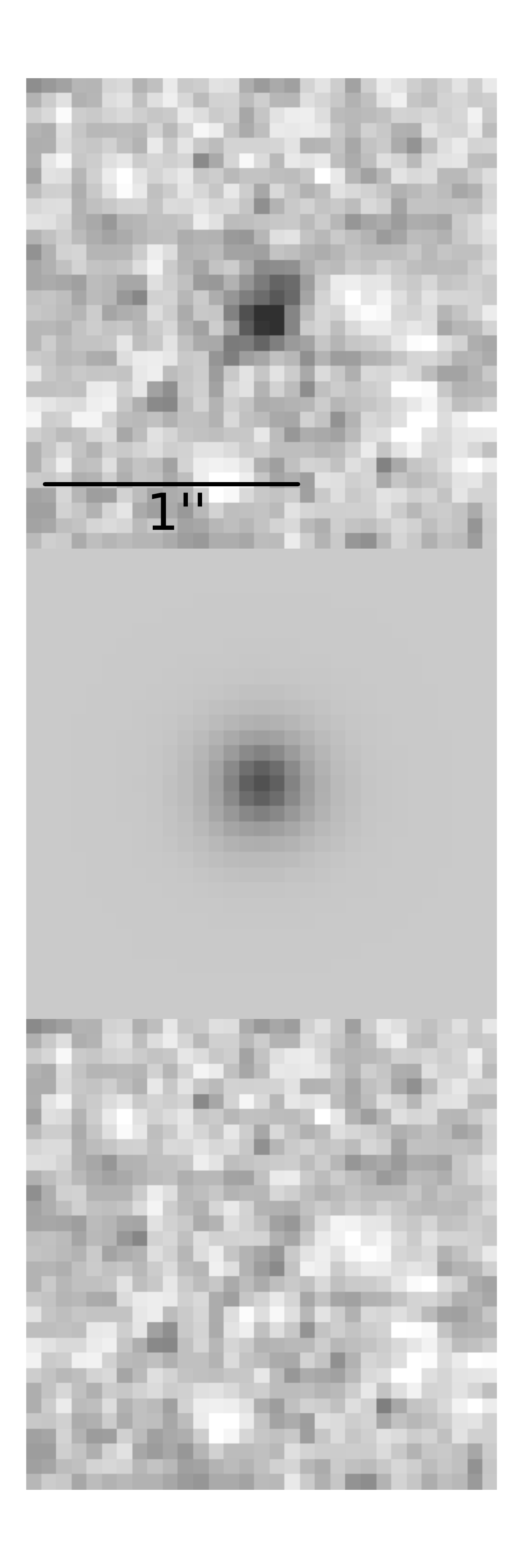}
\end{center}
\end{minipage}
\begin{minipage}{0.11\linewidth}
\begin{center}
% XDFj-38116243\\
\includegraphics[width=\textwidth]{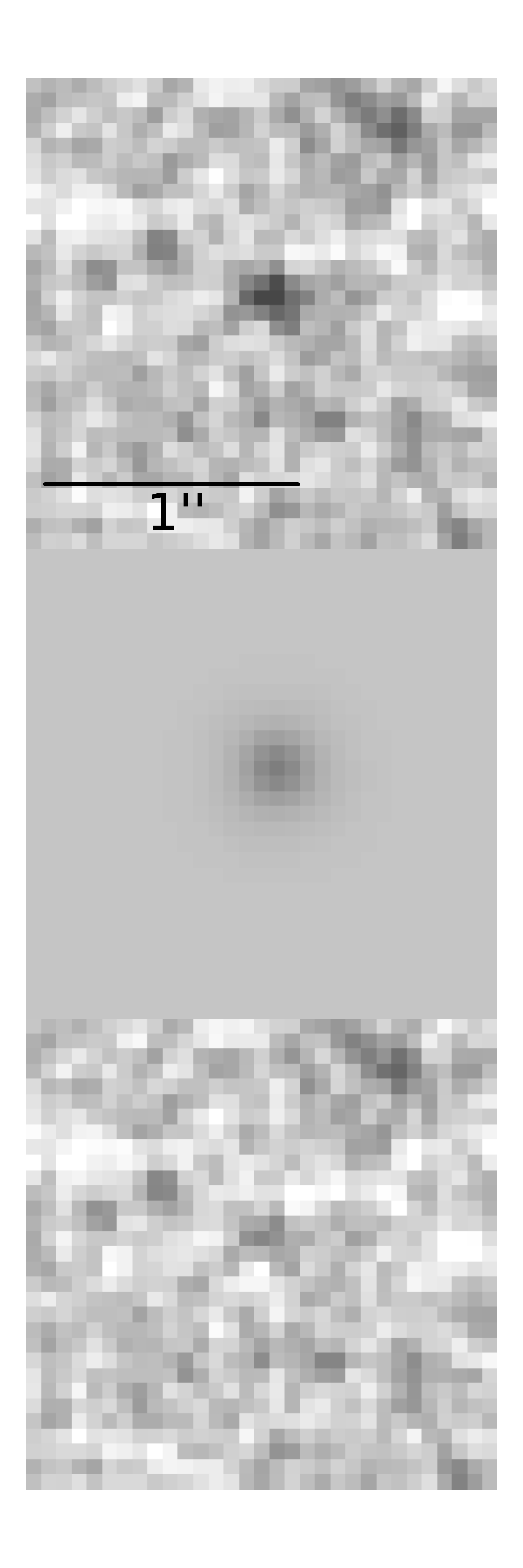}
\end{center}
\end{minipage}
\begin{minipage}{0.11\linewidth}
\begin{center}
% XDFyj-40248004\\
\includegraphics[width=\textwidth]{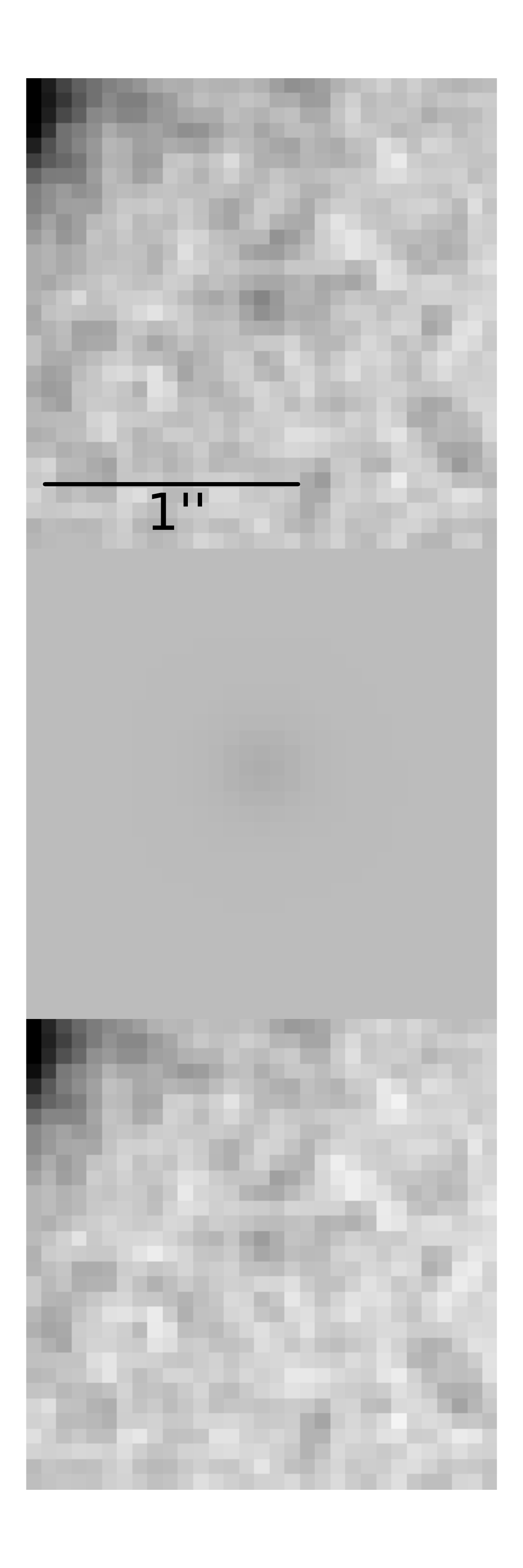}
\end{center}
\end{minipage}\hfill
\caption{The object {\em F160W} cutouts (top row), galfit models (middle row) and residual image (bottom row) for the eight $z\sim9-10$ candidate galaxies we consider here, ranked by apparent luminosity from bright to faint (Table \ref{t:sample}, Figure \ref{f:overview}). Grayscale is $-2\sigma$ to $7\sigma$ centered on the background level and the scale-bar is 1". The six leftmost objects are the bright candidates identified by Oesch et al. (2014) in the GOODS North (GN) and GOODS South (GS) CANDELS fields, with the faint candidates identified in the XDF on the right.}
\label{f:stamps}
\end{center}
\end{figure*}

\begin{deluxetable*}{l l l l l l l l l l l}
 \tablecolumns{10} 
 \tablewidth{0pc} 
\tablecaption{\label{t:sample} The z $\sim$ 9-10 candidates from the XDF and CANDELS fields from \protect\cite{Oesch14} and \protect\cite{Bouwens14}.}
\tablehead{
\colhead{Object ID}		& \colhead{RA}		& \colhead{Dec	}	& \colhead{$H_{160}$}	& \colhead{$z_{phot}$}			& \colhead{$r_{e}$} 				& \colhead{$S/N$}	& \colhead{$log_{10}(M^*)$} 	& \colhead{$L/L^*_{z=3}$}	&  \colhead{$\chi^2$}	& \colhead{References\tablenotemark{g}} \\
 \colhead{}			& \colhead{(J2000)}	& \colhead{(J2000)}	& \colhead{}			&   \colhead{(\tablenotemark{a})}	& \colhead{(kpc)\tablenotemark{b,c}}	& \colhead{(\tablenotemark{d})}				&  \colhead{($M_\odot$)\tablenotemark{e}}	&  \colhead{}	&  \colhead{}	&  \colhead{}		}
\startdata
GN-z10-1 			& 12:36:25.45 & +62:14:31.6 	 & 25.95$\pm$0.07  & 10.2 $\pm0.4$ 		&  0.6$\pm$ 0.3 				& 18.1  	& 9.36  & 1.57 &  18.70  	& [1], [2]\\ 
GS-z9-1 			& 03:32:32.05 & -27:50:41.7 	 & 26.60$\pm$0.20  &  9.3 $\pm0.5$ 		&  0.8$\pm$ 0.2\tablenotemark{f} 	& 5.7  	& 9.17  & 0.76 &  1.21  	& [1]  \\ 
GN-z9-1 			& 12:36:52.24 & +62:18:42.7 	 & 26.62$\pm$0.14  &  9.2 $\pm0.3$ 		&  0.6$\pm$ 0.1 				& 9.0  	& 9.20  & 0.73 &  1.28  	& [1]  \\ 
GN-z10-3 			& 12:36:04.08 & +62:14:29.9 	 & 26.76$\pm$0.15  &  9.5 $\pm0.4$ 		&  0.4$\pm$ 0.1\tablenotemark{*} 	& 9.0  	& 9.17  & 0.67 &  1.38  	& [1], [2] \\ 
GS-z10-1 			& 03:32:26.97 & -27:46:28.3 	 & 26.90$\pm$0.20  &  9.9 $\pm0.5$ 		&  0.5$\pm$ 0.1 				& 7.2  	& 9.10  & 0.63 &  2.00  	& [1], [2] \\ 
GN-z10-2 			& 12:37:22.73 & +62:14:22.7 	 & 26.81$\pm$0.14  &  9.9 $\pm0.3$ 		&  0.5$\pm$ 0.1 				& 7.8 	& 9.15  & 0.68 &  1.34 	& [1], [2] \\ 
XDFj-38116243 	& 03:32:38.11 & -27:46:24.3 	 & 29.87$\pm$0.40  &  9.9 $^{+0.7}_{-0.6}$ 	&  0.3$\pm$ 0.1\tablenotemark{*} 	& 4.7 	& 8.06  & 0.04 &  1.48  	& [2], [3], [4] \\ 
XDFyj-40248004 	& 03:32:40.23 & -27:48:00.3 	 & 29.87$\pm$0.30  &  8.9 $^{+0.6}_{-0.3}$ 	&  0.5$\pm$ 0.9 				& 4.3 	& 7.63  & 0.04 &  5.38  	& [2],[3] 
\enddata
	\tablenotetext{*}{Indicates a marginally resolved source.}
	\tablenotetext{a}{Photometric redshifts from \protect\cite{Oesch14} for the GN sources (their Table 2) and \protect\cite{Bouwens14}, using the {\sc ZEBRA} photometric redshift code \protect\citep{ZEBRA} in both cases.}
	\tablenotetext{b}{Median uncertainty in these $r_e$ values is 0.2 kpc.}
	\tablenotetext{c}{This is the major axis size reported by  {\sc galfit} with the axis ratio fixed such that q=1; equivalent to the circularized radii $r=\sqrt{q}\times r_e$ found elsewhere.}
	%\tablenotetext{d}{The signal-to-noise calculated from the {\sc galfit} output:  ${ F/(\pi*0\farcs35^2) \over \sigma_{sky}}$ with $F$ the flux in {\em F160W} in a $0\farcs35$ aperture and $\sigma_{sky}$ the mean pixel-to-pixel rms noise in the drizzle weight image. We can adopt this definition as the CANDELS data is sufficiently finely drizzled that the pixels are effectively un-correlated. We chose a 0\farcs35 aperture conform with the estimate from \protect\cite{Oesch13a} and \protect\cite{Oesch14}.}
%	\tablenotetext{d}{The signal-to-noise calculated from the {\sc galfit} output:  ${ F/(\pi*0\farcs18^2) \over \sigma_{sky}}$ where $F$ is the flux in {\em F160W} in a $0\farcs18$ radius aperture and $\sigma_{sky}$ mean pixel-to-pixel rms noise from the drizzle weight image. We chose a 0\farcs18 aperture radius to conform with the estimate from \protect\cite{Oesch13a} and \protect\cite{Oesch14}.}
	\tablenotetext{d}{The signal-to-noise calculated from the light enclosed in a 0\farcs36-diameter aperture \citep[see][]{Bouwens14}, conform with the estimate from \protect\cite{Oesch13a} and \protect\cite{Oesch14}.}
	\tablenotetext{e}{Mass estimates are from \cite{Oesch14} (their Table 6). Mass estimates for $z\sim9-10$ candidates from the XDF data assume the same mean $M/L_{F160W}$ ratio, i.e., $0.32 ~ M_\odot/L_\odot$ as \protect\cite{Oesch14} found for their bright sources. }
	\tablenotetext{f}{Galfit parameters for GS-z9-1 are available from the CANDLES team \citep[][
	\url{http://www.mpia-hd.mpg.de/homes/vdwel/candels.html}]{van-der-Wel12,van-der-Wel14}, which lists: m=27.1$\pm$0.3, $r_{e}=0\farcs05\pm0.03$ (corresponding to 0.2 kpc) and n=2.83$\pm$2.81 (flag=2, S/N=6.1). The difference in $r_e$ can be attributed to \cite{van-der-Wel14}'s leaving the S\'{e}rsic index free and differences in our segmentation maps for the source. For more details, see \S \ref{s:prevresults}. While van der Wel et al. (2014) find a smaller size for this source, overall our size measurements agree quite well (within $\sim$20\%) with those from \protect\cite{van-der-Wel14} and \protect\cite{Grazian12}.}\\
	\tablenotetext{g}{References: 
	[1] \protect\cite{Oesch14}, 
	[2] \protect\cite{Bouwens14}, 
	[3] \protect\cite{Oesch13}, 
	[4] \protect\cite{Bouwens11b}  } 		% GOODS-N
\end{deluxetable*}%
\section{Observational Data}
\label{s:data}

To measure the sizes, we use the public data from the XDF \citep{XDF} and CANDELS  \citep{Koekemoer11, Grogin11} fields. The size measurements are performed in the {\em F160W} filter drizzled images from both programs. Pixel scales are set to 0\farcs06 (compared to the native 0\farcs13 for WFC3/IR) and $5\sigma$-limiting {\em F160W} magnitudes are 29.8 (XDF), 28.4 (CANDELS-deep) and 27.6 (CANDELS-wide) respectively for a 0\farcs35-diameter aperture. 

\section{Methodology for Size Measurements}
\label{s:method}

A convenient and powerful tool to measure sizes accurately for faint sources is {\sc galfit} \citep{galfit,galfit2}.
{\sc galfit} determines the size of an object by comparing the two-dimensional profile of a galaxy with a 
PSF-smoothed S\'{e}rsic profile and then finding the model which minimizes the value of $\chi^2$. 
We fix the S\'{e}rsic index to $n=1.5$ in our fits (see Table \ref{t:galfit}, consistent with the S\'{e}rsic 
parameters derived for stacked $z=4-6$ samples in \cite{Hathi08}. Fixing the Sersic index to other values 
(i.e., n=1-2.5) did not change the effective radius result significantly ($<10$\%).
We allow the central position to range within 3 pixels of the one determined by {\sc sextractor} 
({\sc x\_peak, y\_peak}). In the case of a single object (XDFyj-40248004), the center was fixed to the 
{\sc sextractor} value. We use {\sc sextractor} to estimate the local background (128 pixels aperture) 
and the drizzle weight map for an estimate of noise for both the application of {\sc sextractor} and 
{\sc galfit}. The sextractor run on the field yields an initial guess of the position angle and effective 
radius for {\sc galfit} {(see Table }\ref{t:galfit}). We fix the axes ratio ($q=1$) as these objects are mostly circular. We did try fits with {\sc galfit} with the axes ratio ($q$) as a free parameter but the resulting axes ratio was too uncertain to be informative. 

While our CANDELS and XDF reductions are already globally background-subtracted, we
estimate the local backgroundÊsurrounding the fit objects again with {\sc GALFIT} to
ensure that local variations do not influence the fit results.

The dominant uncertainties in the measured sizes are the estimated background, the precise shape of the 
point spread function (PSF), and the pixels included in the fit. For an in-depth discussion on the uncertainties 
and biases in size measurements with {\sc galfit} we refer to \cite{Ono13}.

% PSF
We use repeat fits of each object to estimate variance due to different PSF models.
These models are from the 3D-HST project \citep[v3.0][\url{http://3dhst.research.yale.edu}]{Brammer12,van-Dokkum13a, Skelton14} each
derived for a specific CANDELS field, resulting in unique outer structure, and an additional HST PSF with forced circular symmetry.
Similar to \cite{van-der-Wel14}, we find that the choice of PSF model only has a 
minor impact on the effective radius measurement, i.e., fit-to-fit variance is much lower than the error.

%SEGMENTATION
An important input value for {\sc galfit} is the list of pixels to include in the shape fit. One can use either those pixels
attributed to an object by {\sc sextractor} (segmentation map), all pixels in an image {\em except} those assigned to 
other objects (masked) or simply all the pixels in a cut-out area. This latter option is preferred for faint and isolated 
sources to minimize bias and we opt for this as the fitted sources are generally isolated from neighboring objects (Figure \ref{f:stamps}).
In the case of GN-z10-1 and to a lesser degree GS-z10-1 and XDFyj- 40248004, there are other sources in the {\sc galfit} stamp.
However, we found that our fit results did not improve appreciably when masking neighboringÊobjects. 
We found that it was sufficient to limitÊthe central pixel position soÊthat it was closeÊtoÊthe one found by {\sc SExtractor}.

\begin{table*}
\caption{{Galfit settings and parameters. Parameters marked with (se) are script variables for which we used the }{\sc sextractor} values.}
\begin{center}
\begin{tabular}{l l l}
% 	& IMAGE and GALFIT CONTROL PARAMETERS & \\
\hline
\hline
Order	& Default, example or script variable		&	Galfit description \\
A) 	& 	GOODS-S\_F160W\_stamp.fits			&	 Input data image (FITS file)\\
B) 	& 	object\_name\_model.fits      			& 	 Output data image block\\
C) 	& 	rms.fits            						&	 Sigma image name (made from data if blank or "none") \\
D) 	& 	GOODS-S\_F160W\_psf.fits          		& 	 Input PSF image and (optional) diffusion kernel\\
E) 	&	1    					 	 	  	&	 PSF fine sampling factor relative to data \\
F) 	& 	seg.fits           						& 	 Bad pixel mask (FITS image or ASCII coord list)\\
G) 	& 	none                						&	 File with parameter constraints (ASCII file) \\
H) 	& 	$x_c$(se)-50 $x_c$(se)+50 $y_c$(se)-50 $y_c$(se)+50	&	 Image region to fit (xmin xmax ymin ymax)\\
I) 	&	180    180          						&	 Size of the convolution box (x y)\\
J) 	& 	25.9463		 				 	&	 Magnitude photometric zeropoint \\
K) 	&	0.060  0.060       					&	 Plate scale (dx dy)   [arcsec per pixel]\\
O) 	&	regular             						&	 Display type (regular, curses, both)\\
P) 	&	0                   						&	 Choose: 0=optimize, 1=model, 2=imgblock, 3=subcomps\\
\hline
 0) 	&	sersic                 						&	  Component type\\
 1) 	&	$x_c$(se) $y_c$(se) 1 1  				&	  Position x, y\\
 3) 	&	$m_{160}$(se)     1          				&	  Integrated magnitude \\
 4) 	&	$r_e$(se)      1          					&	  $R_e$ (effective radius)   [pix]\\
 5) 	&	1.5000      0         					&	   Sersic index n (de Vaucouleurs n=4) \\
 6) 	&	0.0000      0          					&	     ----- \\
 7) 	&	0.0000      0         					&	      ----- \\
 8) 	&	0.0000      0          					&	     ----- \\
 9) 	&	1.0		0          					&	  Axis ratio (b/a)  \\
10) 	&	PA(se)     1          					&	  Position angle (PA) [deg: Up=0, Left=90]\\
 Z) 	&	0                      						&	  Skip this model in output image?  (yes=1, no=0)\\
\hline
\hline
\end{tabular}
\end{center}
\label{t:galfit}
\end{table*}%

\section{Results}
\label{s:results}

Here we measure sizes for the six particularly bright $z\sim9-10$ candidates discovered by \cite{Oesch14} in CANDELS and two faint candidates identified by \cite{Oesch13} and \cite{Bouwens14} from the XDF. The bright $z\sim9-10$ candidates were identified using a $J-H > 0.5$, $H - [4.5] < 2$, and  optical+Y-non-detection criterion, while the faint $z\sim10$ candidates were identified with $J-H>1.2$, $H-[3.6]<1.4$, and optical+Y-non-detection criterion.

Figure \ref{f:stamps} shows the {\em F160W} data, our {\sc galfit} model and the residual image for two $z\sim9-10$ candidate galaxies. The reported values are the $R_e$ value from galfit, i.e. the  effective radius along the major axis but with the axes ratio ($q$) fixed to unity and therefore identical to the ``circularized'' radius ($\sqrt{q}\times R_e$). Typical half-light radii are between 0\farcs10 and 0\farcs25, corresponding to $\sim0.5$ kpc at their respective redshifts (Table \ref{t:sample}). The mean uncertainty in effective radius is 0\farcs06 (0.28 kpc). Fits to these faint objects are reasonably good (reduced $\chi^2 \sim 1-19$). Table \ref{t:sample} lists the $H_{160}$ apparent magnitude determined with {\sc galfit} and the luminosity and inferred stellar mass from that value.

\begin{figure}
\begin{center}
% ../../Wyithe_z_reff.pdf
\includegraphics[width=0.5\textwidth]{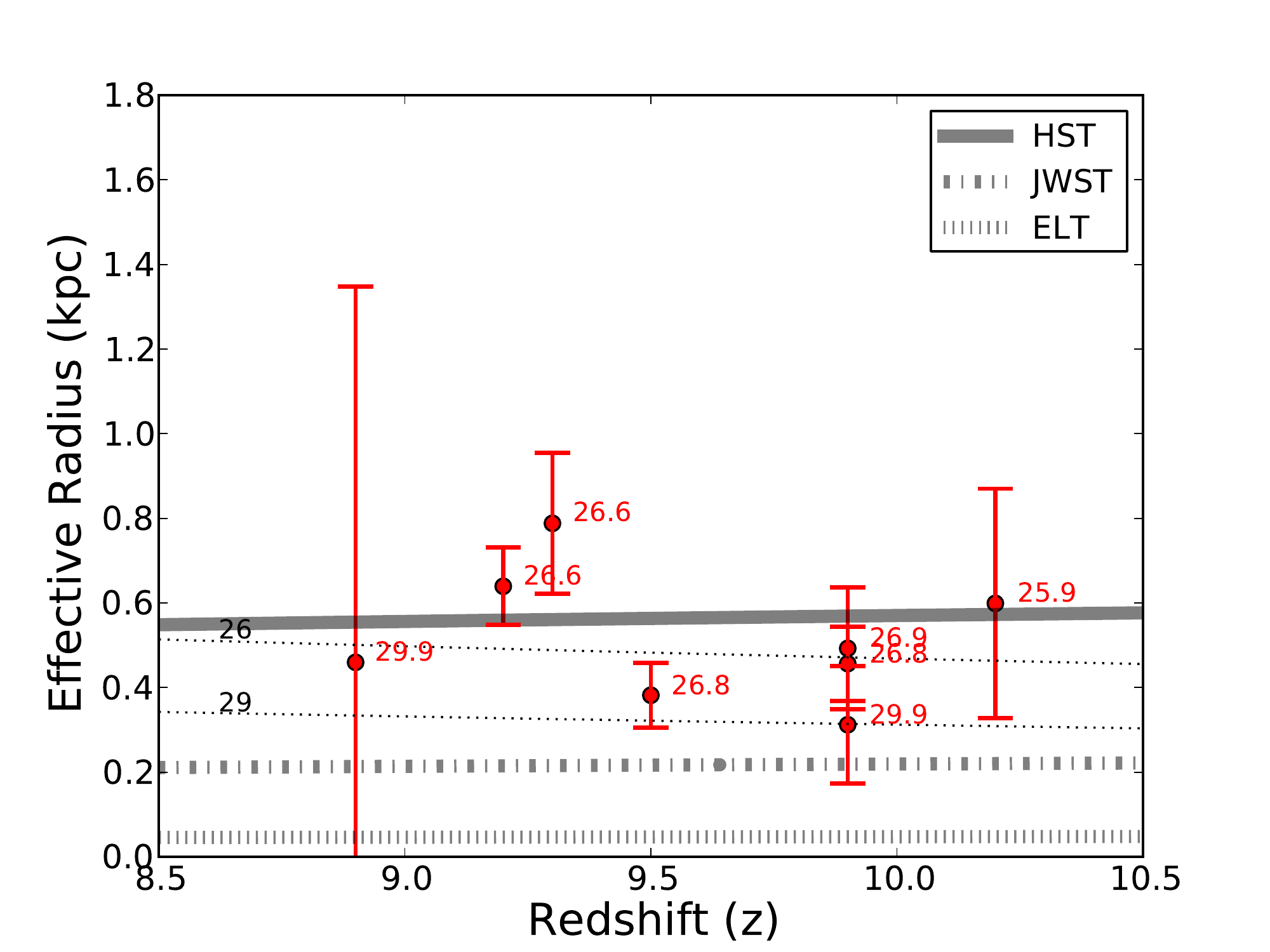}
\caption{The expected size-redshift relation from the simple model of \cite{Wyithe11} (equation \ref{eq:model}), the nominal diffraction limits for HST/WFC3, JWST and a 30m ELT (equation \ref{eq:telescope}) and our size measurements at their photometric redshifts with their apparent AB magnitude (red text). The thin dotted lines are based on the \cite{Wyithe11} model for galaxy sizes (equation \ref{eq:model}) with an $\alpha=-2.27$ slope for the luminosity function. This value is the determination from \protect\cite{Bouwens14} for the $z\sim10$ population. The agreement between the data and the model anchored on earlier observations illustrates that while a simple model suffices to predict sizes, it needs to be anchored to high-redshift measurements if one is to plan observations of these earliest epochs of galaxy evolution with future observatories such as JWST and ELT.}
\label{f:model}
\end{center}
\end{figure}

\cite{Ryan11} and \cite{Holwerda14} explore the {\sc sextractor} effective radius ($r_e$) of known Galactic stars in CANDELS. They consider sources with $r_e < 0\farcs15$ (uncorrected for the PSF)  to be unresolved (0\farcs1 corrected for the {\em F125W} PSF). In the case of {\sc galfit}, the minimum effective radius can be smaller because the model is convolved with the PSF. Figure \ref{f:model} illustrates how the majority of our sources are indeed resolved with HST.Ê

Two of the candidate high-redshift galaxies have {\sc galfit} effective radii indicating they are marginally resolved sources ($r_{e} < 0\farcs1$), one from CANDELS (GN-z10-3) and one in the XDF (XDFj-38116243). The CANDELS sample is therefore better resolved compared to the XDF sources: 50\% compared to 17\%. The mean effective radii are $<r_e>=0\farcs09$ (XDF) and $<r_e>=0\farcs13$ (CANDELS), respectively, illustrating the benefits of the latter sample.

\subsection{Comparison to previous results}
\label{s:prevresults}

As a check on the {\sc galfit} sizes, we measure the sizes for $z\sim7$ galaxy candidates from the CANDELS South \citep{Bouwens14} in the same manner as the $z\sim9-10$ candidates and compare these to the results from \cite{van-der-Wel14}. Both size measurements were obtained from the CANDELS {\em F160W} mosaics. We note however, that we made use of the reductions from the 3D-HST team \citep{Skelton14} and an RMS map based on the drizzle weight map \citep[see][]{Casertano00, seman}. We find good agreement in the mean between the {\sc galfit} radii we measure and the size measurements in the \cite{van-der-Wel14} catalog (within 14\% for 29 sources). We note that \cite{van-der-Wel14} leave the S\'{e}rsic index as a free parameter, ranging up to $n=3$, whereas we keep it fixed ($n=1.5$). Refitting the same 29 $z\sim7$ sources with the \cite{van-der-Wel14} reported S\'{e}rsic indices ($n$), we arrive at similar sizes ($< 0\farcs05$ difference).

A second check is provided by the $z\sim7$ {\sc sextractor} catalog from \cite{Grazian12}, and assuming exponential disks at this redshift. Matching this catalog against our $z\sim7$ catalog, we find reasonable agreement between the measured sizes (within $\lesssim20$\%) for the eight galaxies present in both samples.

\begin{figure}
\begin{center}
% Reff_arcsec_IRAC_H_36_3panel_v2.pdf
\includegraphics[width=0.5\textwidth]{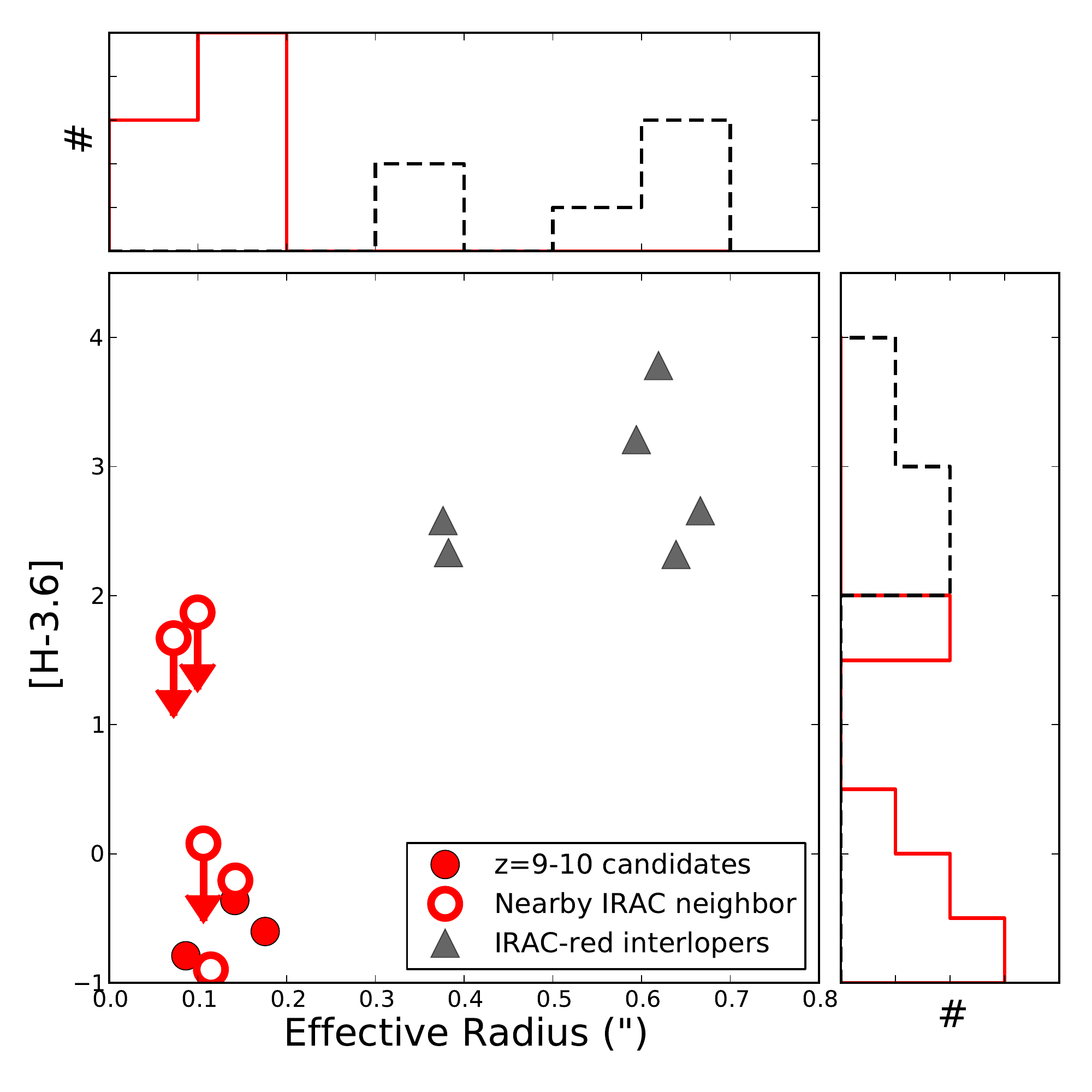}
\caption{The $z\sim9-10$ galaxy candidates from  \cite{Oesch13a} and \cite{Oesch14} and the interlopers to the $z\sim9-10$ selection where the $H-[3.6]$ or $H-[4.5]$ criterion is not applied. Those objects withÊclose companions in the IRAC 3.6 $\mu$m images (see Figure 4 in \protect\cite{Oesch13a} and Figure 2 in \protect\cite{Oesch14}) areÊindicatedÊwith open circles and with arrows where lower limits on the flux measurements are relevant. The effective radii ($r_e$) and  $H-3.6$ colors for $z\sim9-10$ galaxy candidates from \protect\citet[][red circles]{Oesch14} and interlopers to a $z\sim9-10$ selection (gray triangles) when no $H - [3.6] $ or  $H - [4.5]$ criterion is applied. The $z\sim9-10$ galaxies and the lower-redshift interlopers separate well in both their measured $H - [3.6]$ colors and sizes. This suggests that the sizes of $z\sim9-10$ candidates could serve as an alternate constraint on the high-redshift nature of $z\sim9-10$ candidates where no Spitzer/IRAC data are available (e.g., as with the BORG program).}
\label{f:conf}
\end{center}
\end{figure}

\subsection{Confirmation of \cite{Oesch14} Photometric $z\sim9-10$ Selection}

It is useful to compare the sizes we measure for the bright $z\sim9-10$ candidates from \cite{Oesch14} with that expected extrapolating the sizes of lower-redshift galaxies to $z\sim10$ to see if the candidates seem consistent with lying at $z\sim10$ as the strong photometric evidence would suggest. The mean effective radius for both bright and faint sources ($<r_e>=0.6$ and 0.4 kpc, respectively) conforms to general expectations for star-forming galaxies at $z\sim9-10$ (see \S 4.6). For comparison, we also measure {\sc galfit} sizes for six potential interlopers to the \cite{Oesch14} selection, i.e., satisfying all the $z\sim9-10$ criteria except that their H$-$[3.6] or H$-$[4.5] color is very red. These low-redshift and likely dusty interlopers have a mean effective radius of $<r_e>=0\farcs59$, substantially larger than the $<r_e>\sim0\farcs13$ radius we find for our bright $z\sim9-10$ sample. 

In spite of the encouraging results from this test, we emphasize that the strongest constraints on the high-redshift nature of these sources come from the accurate photometric observations available for each of these sources, which combined strongly favor a $z\sim9-10$ identification.  Nevertheless, this size test does provide evidence that the \cite{Oesch14} candidate $z=9-10$ galaxies do not correspond to some exotic (yet unseen) population of contaminants, as appears to have occurred when an unprecedented extreme emission-line galaxy (EELG) contaminated the \cite{Bouwens11b} $z=10$ photometric selection.

More generally, we have found a general correlation between the measured size of sources with {\em HST} photometry consistent with their corresponding to $z\sim9-10$ galaxies and their $H - \textrm{IRAC}$ colors.  As shown in Figure \ref{f:conf}, the measured sizes of \cite{Oesch14} candidates are all smaller than any of the IRAC-red interlopers to the  \cite{Oesch14} selection. 
We note that the IRAC-red interlopers are about equally numerous as sources in our $z\sim9-10$ sample. Without information from deep Spitzer/IRAC imaging on these sources, they could constitute a serious contamination of any $z>9$ sample. Size could therefore serve as a proxy for IRAC color information, where the latter is lacking. The separation in both H-[3.6] color and size in Figure \ref{f:conf} adds confidence that our candidate $z=9-10$ galaxies are indeed at these redshifts.

\subsection{Size-Luminosity Relation}
\label{s:Mabs-r}

\begin{figure}
\begin{center}
\includegraphics[width=0.5\textwidth]{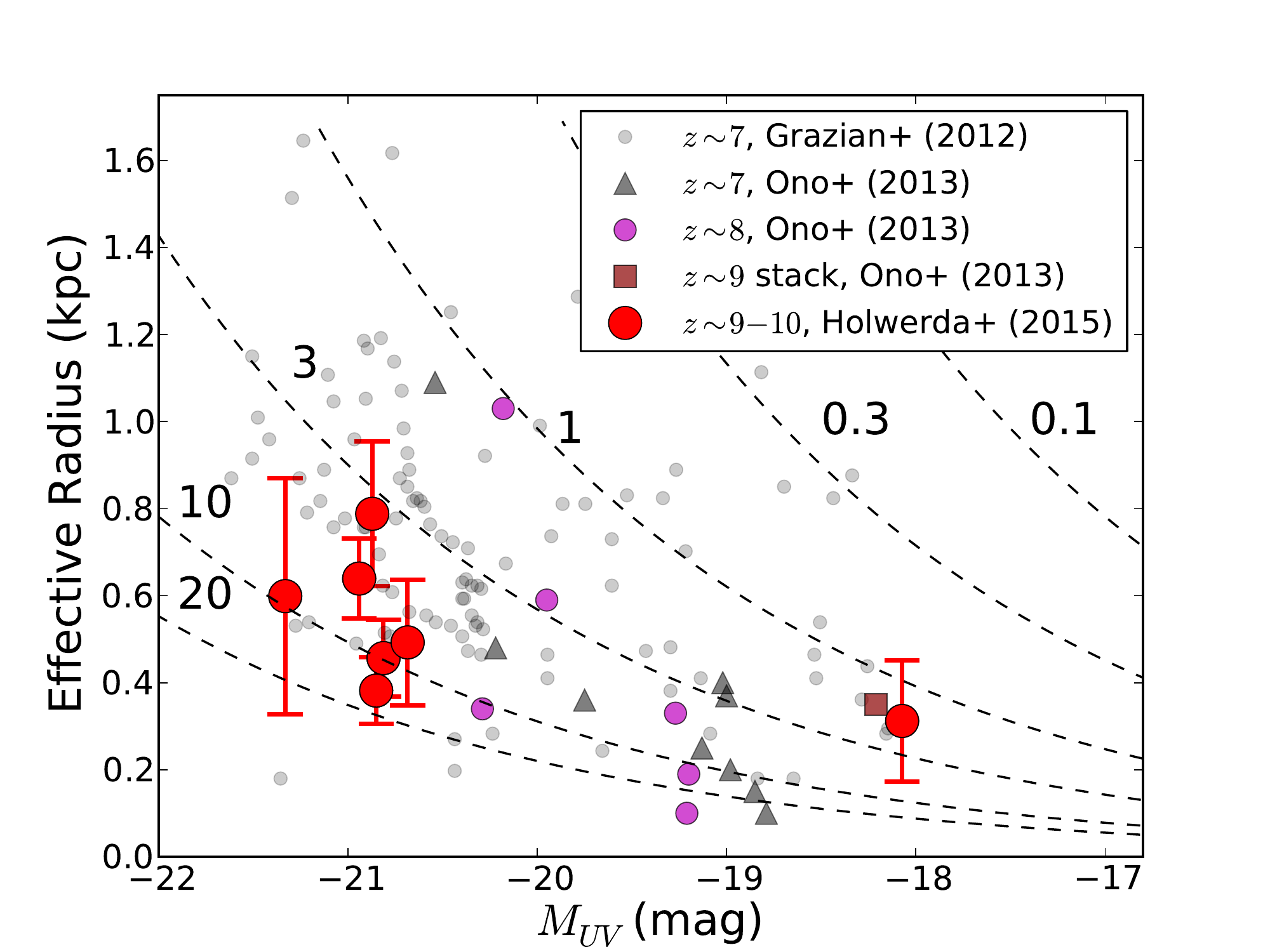}
\caption{Observed UV luminosities (1450 \AA~  at z=10) vs. effective radius for our sample of $z\sim9-10$ candidate galaxies (excluding the galaxy with the most uncertain size, XDFyj-40248004). Light gray points are the sizes and absolute luminosities in {\em F160W} from \protect\citet[][{PSF-corrected}]{Grazian12}, for their $z\sim7$ sample, while the dark grey, magenta and brown points are the $z\sim7$, $z\sim8$, and $z\sim9-10$ samples respectively, from \protect\cite{Ono13}. The red dotted line and shaded area are the best fit and $1\sigma$ uncertainty on the luminosity-size relation for our sources. The dashed lines are star formation surface density levels of $\rm \Sigma_{SFR}=0.1, 0.3, 1, 3, 10, 20  ~ M_\odot~ yr^{-1}~ kpc^{-2}$ (assuming no dust or emission lines in the rest-frame UV).}
\label{f:SFR}
\end{center}
\end{figure}

The observed correlation between the physical sizes of galaxies and their luminosities provides us with information on how the physical scale of star-forming regions in galaxies scales with the SFR across cosmic time.  Figure \ref{f:SFR} shows the relationship between the UV luminosity and the effective radius for our $z=9$-10 sample, with the exception of one source from our faint sample XDFyj-40248004 which is not shown, since its size measurement is quite uncertain (Table \ref{t:sample}). For comparison, we also include the sizes from \cite{Grazian12} and  \cite{Ono13} in this Figure. The caveats are that (1) UV is sensitive to longer-duration star-formation, (2) the conversion from UV luminosity to SFR for low-metallicity, high-mass stars is uncertain,Êand (3) a lack of  emission line observations, which track more stochastic star-formation. Our candidate $z\sim9-10$ galaxies have a similar range of sizes and luminosities as the $z\sim7-8$ galaxies from \cite{Grazian12} and  \cite{Ono13}. 

To help interpret the relationship between size and luminosity, it is useful to fit a linear relation to the points in Figure \ref{f:SFR}:
\begin{equation}
R = R_* \left({ L \over L_*(z=3) }\right)^b
\end{equation}
\noindent similar to the treatment at lower redshifts \citep{Huang13}. The slopes ($b$) we find for the $z=9$-10 sample, those from  \cite{Huang13}, and the ones we derive for the \cite{Mosleh12} and \cite{Ono13} samples are presented in Table \ref{t:Lslopes}. The slopes of the luminosity-size relation for \cite{Ono13} and this work are very uncertain due to the small number of sources in the present samples, but are plausibly consistent with what has been found at lower redshifts.  All values of the luminosity-slope relation are consistent with a $b\sim0.25$ slope for the entire redshift range.

The outstanding issue with the $z\sim9-10$ size-luminosity relation is that the statistical weight for the fit of the slope hinges on a single point (XDFj-38116243). However, we note that the stacked $z\sim9-10$ result from \cite{Ono13} is near this point as well. Not only will a large sample size improve the determination of the relation, preferably, it will need to be distributed over more than a magnitude of luminosity (Figure \ref{f:SFR}), as well as include a correction for strong emission lines.

\begin{table}
\caption{The slopes of the luminosity-size relation (top ones from the summary in \protect\cite{Huang13}). 
The slopes for the \protect\cite{Mosleh12}, \protect\cite{Grazian12} and \protect\cite{Ono13} samples were derived by us based on the published values.}
\begin{center}
\begin{tabular}{l l l l}
Redshift	& Intercept			& Slope 		& Reference\\
(z)		& ($R_*$, kpc)				& ($b$) 	& \\
\hline
\hline
0		& 5.93 $\pm$ 0.28			& 0.21 $\pm$ 0.03		& (1)\\
0		& 3.47					& 0.26 				& (2)\\
0 		& --						& 0.32 $\pm$ 0.01		& (3)\\
4		& $1.34^{+0.10}_{-0.11}$		& 0.22 $\pm$ 0.06		& (4)\\
5		& $1.19^{+0.21}_{-0.16}$		& 0.25 $\pm$ 0.15		& (4)\\
\hline
7  		& 0.86 $ \pm $ 0.04 			& 0.24 $ \pm $ 0.06 		& (5) \\
7		& 1.55 $ \pm $ 0.31 			& 1.05 $ \pm $ 0.21 		& (6) \\
8		& 1.44 $ \pm $ 1.07  			& 1.03 $ \pm $ 0.75 		& (6) \\ 
9-10		& 0.57 $ \pm $ 0.06  			& 0.12 $ \pm $ 0.09 		& This work.\\
\hline
\end{tabular}
\end{center}
(1)  \cite{de-Jong00} \\
(2) \cite{Shen03} \\
(3) \cite{Courteau07b} \\
(4) \cite{Huang13} \\
(5) derived by us from the \cite{Grazian12} data\\
(6) derived by us from the \cite{Ono13} data\\
\label{t:Lslopes}
\end{table}

\subsection{Star-formation Rate Surface Density}

The sizes and absolute luminosity in the rest-frame ultra-violet are intimately linked to the SFR density in these systems, informing us of the conditions in these first stellar systems.
The SFR surface density can be tied to the absolute UV magnitude and effective radius by 
\begin{equation}
\rm M_{UV} =-2.5 \times log_{10}\left({ \Sigma_{SFR}\times r_{e}^2 \over  10[pc ~ in ~ cm]^2 \times 2.8\times10^{-28} }\right) -48.6 
\end{equation}
\noindent from \cite{Ono13}, where $\Sigma_{SFR}$ is in $M_\odot~ yr^{-1}~ kpc^{-2}$ and $r_e$ in kpc. Neither dust extinction or strong emission lines are assumed in this conversion.

Figure \ref{f:SFR} shows the relation between our effective radii and the implied absolute magnitudes for different values of the star formation rate surface density. Galaxies in our sample are consistent with $\rm \Sigma_{SFR}\sim1-20 ~ M_\odot~ yr^{-1}~ kpc^{-2}$. \cite{Ono13} found similar SFR surface densities for $z\sim7-8$ galaxies, and \cite{Oesch10b} and \cite{Shibuya15} show there is limited evolution in the average SFR surface density (for  $>0.3L^*_{z=3}$ galaxies) from $z\sim4$ to 8 with their mean bracketed by our values. 

\subsection{Mass-Size Relation}
\label{s:Mr}

\begin{figure}
\begin{center}
\includegraphics[width=0.5\textwidth]{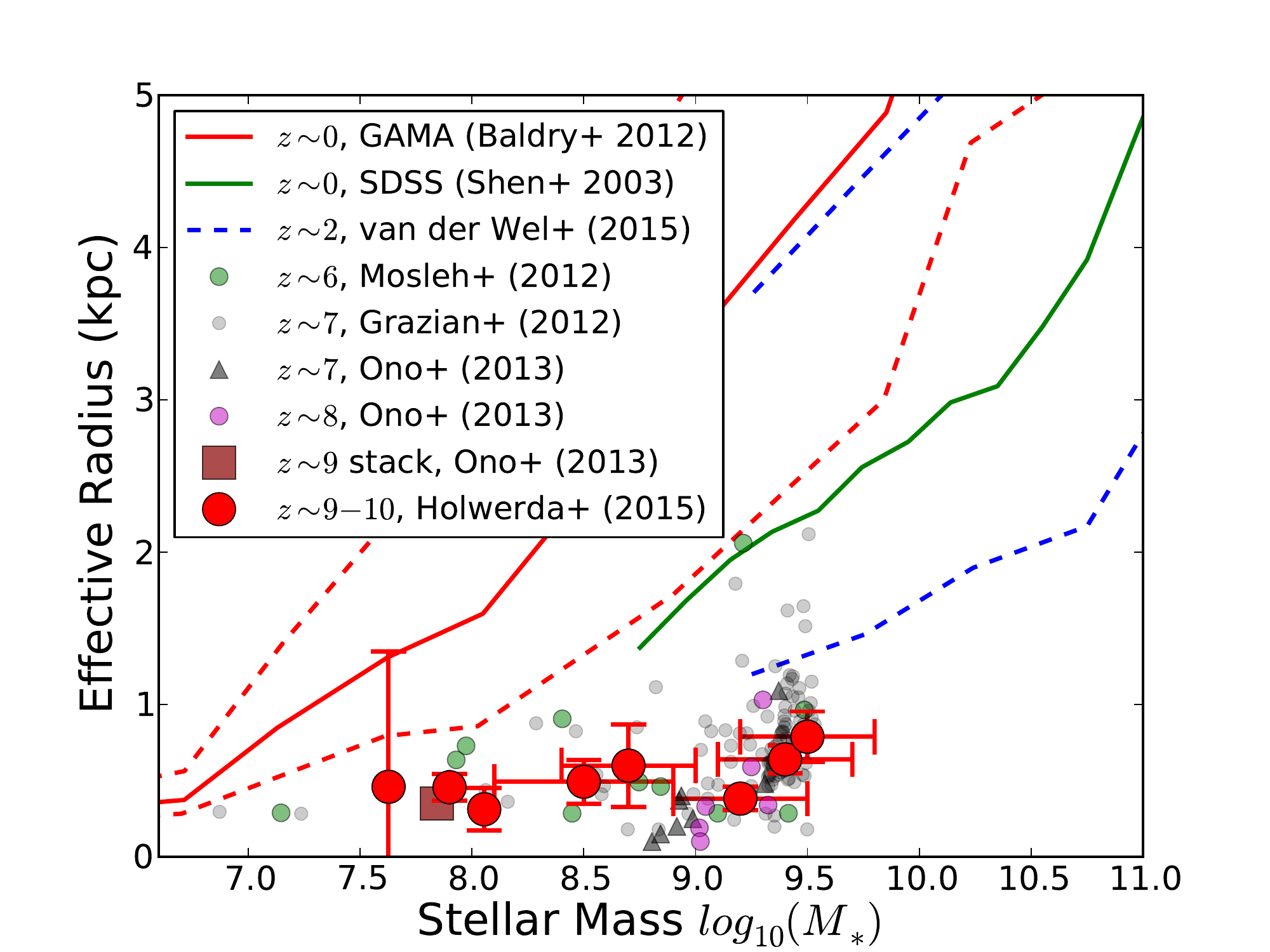}
\caption{The mass-size relationship for our sample of $z\sim9-10$ galaxies from CANDELS and the XDF. For comparison, we show the $z\sim2$ \protect\citep[blue dashed interval,][]{van-der-Wel14}, 
the $z\sim6$ \protect\citep[green points][]{Mosleh12}, the $z\sim7$ and $z\sim8$ galaxies based on the \protect\citep[][light gray]{Grazian12} and \protect\citep[][dark gray and magenta respectively]{Ono13} catalogs \protect\citep[corrected for emission line contamination and adopting a M/L ratio from][ for the latter two respectively]{Stark13}. The dark red square is the single $z\sim9-10$ stacked size measurement from \protect\cite{Ono13}, similarly converted. The mass-size relations for $z=0$ blue galaxies are from GAMA \protect\citep[red line,][]{Baldry12} and SDSS \protect\citep[green line,][]{Shen03}. The thick red dotted line and shaded area are our best fit to the $z\sim9-10$ data with $1\sigma$ uncertainty.}
\label{f:MassReff}
\end{center}
\end{figure}

The availability of size measurements and mass estimates for our sources allows us to examine the mass-size relation to $z\sim9-10$. Our mass estimates for the bright sources are from \cite{Oesch14} and for the HUDF/XDF sources from the $H_{F160W}$ using the mean $M/L$ ratio of the  \cite{Oesch14} values (0.36 $M_\odot/L_\odot$). We caution that there are large potential systematic uncertainties in these estimates, due to the likely presence of nebular emission lines of unknown strength in the IRAC fluxes (rest-frame optical) which \cite{Oesch14} use to derive the masses for their $z\sim9-10$ sample.

Figure \ref{f:MassReff} shows the relation between mass and size with comparison samples at high-redshift z=2, 6, and 7 \citep[][respectively]{van-der-Wel14,Mosleh12,Grazian12,Ono13} and the local relations from SDSS \citep{Shen03} and GAMA \citep{Baldry12}. There is only a very weak mass-size relation compared to the steeper relation at z=0 \citep{Shen03,Baldry12} or z=2 \citep{van-der-Wel14}. Our $z=9$-10 sample occupies the same mass-size space as the z=6 sample from \cite{Mosleh12}. Converted to mass following the \cite{Stark13} prescription, the $z\sim7$ samples from \cite{Grazian12} and \cite{Ono13} have similar sizes to the most massive galaxies from our z=9-10 sample, but with a few outliers to $r_e\sim1.5$ kpc. Overall, we find much weaker evolution in the mass-size relation than in the luminosity-size relation. This is not especially surprising given the evolution in the sSFRs (and hence M/L ratios) of galaxies from $z\sim7$ to $z\sim3$ \citep{Stark13,Gonzalez14}.  The M/L ratio evolution largely cancels evolution in the sizes of galaxies at fixed luminosity, resulting in only a weakly evolving size-mass relation.

To quantify the relation between size and mass, we fit a linear relation to the points in Figure \ref{f:MassReff}:
\begin{equation}
R = R_0 \left({ M_* \over M_0 }\right)^\beta
\end{equation}
where we fix $M_0 = 1. \times 10^{9} M_\odot$ since it corresponds to approximately the median stellar mass of our sample.

\begin{figure}
\begin{center}
% ../../Figures/Mass_Reff_LinFits_b.pdf'
\includegraphics[width=0.5\textwidth]{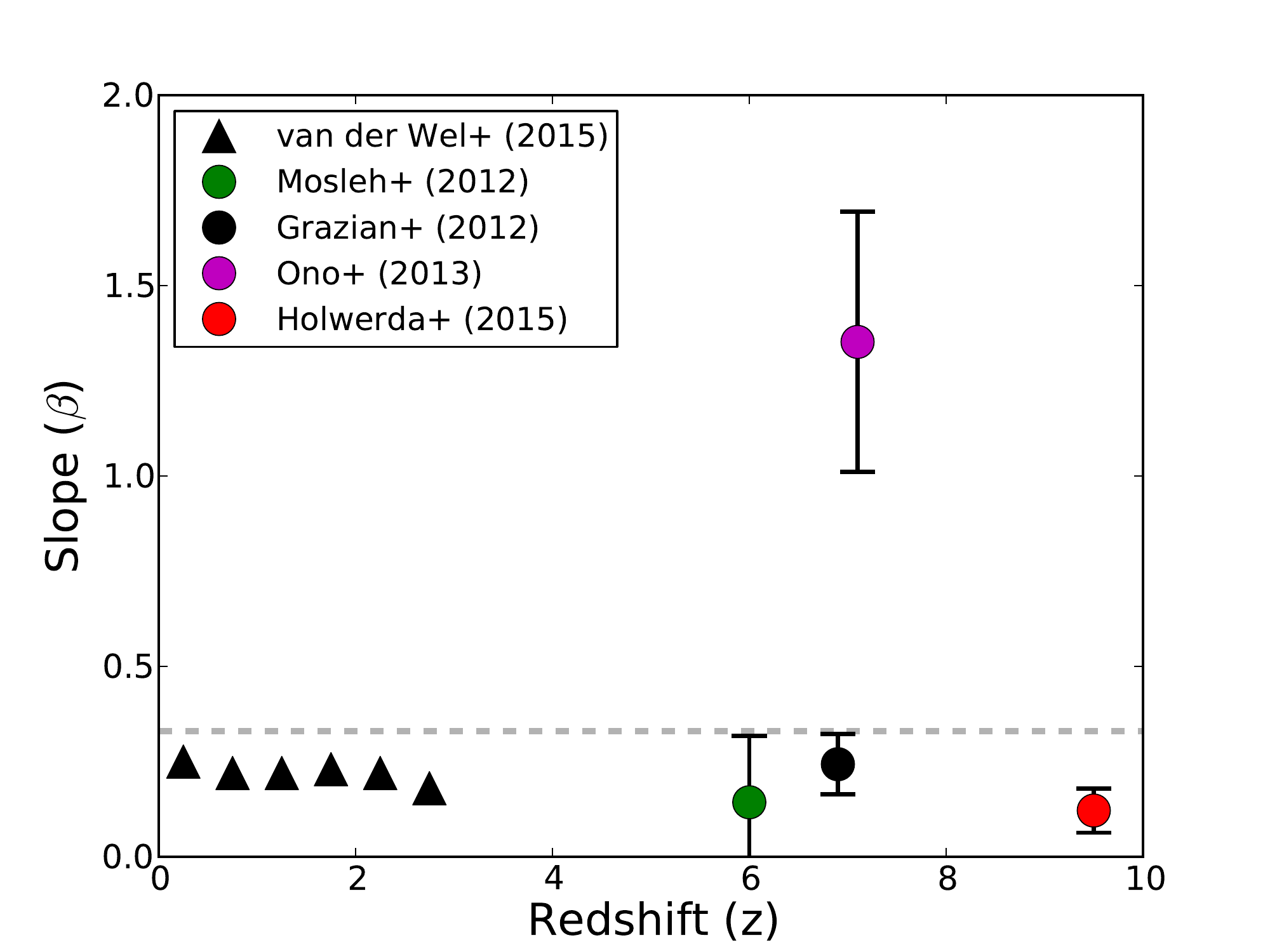}
\caption{The slope of the stellar mass-size relation ($\beta$) as a function of redshift. The lower-redshift points are from \protect\cite{van-der-Wel14}, derived from their fit. Redshift $z\sim6-7$ points are our fits based on the inferred mass-size relations derived from \protect\cite{Mosleh12}, \protect\cite{Grazian12} and \protect\cite{Ono13}, with mass-to-light corrections from \protect\cite{Stark13}. The dashed line is the maximum ($\beta\simeq1/3$). The $z\sim9-10$ sample exhibits a practically flat slope compared to most previous work. }
\label{f:beta}
\end{center}
\end{figure}

The intercept ($R_0$ at $M_0$) and slopes ($\beta$) we find are listed in Table \ref{t:Mslopes} and plotted as a function of redshift in Figure \ref{f:beta}. In general, these slopes are uncertain but comparable with those found for earlier epochs for the luminosity-size relation (of star-forming galaxies), which can be expected if the mass-to-light ratio conversion is not mass-dependent over the range probed. We note that the $z\sim7$ relations based on the \cite{Mosleh12} and \cite{Ono13} samples are poorly constrained.  From theory, the relation between luminosity or stellar mass with size is expected to be slightly shallower than $\beta\sim1/3$ \citep[see e.g.,][]{Dutton12,Stringer14}.

The slope is typically around 0.25 for most star-forming, late-type galaxies over the age of the Universe (see \citet{van-der-Wel14} for mass-size relations, Figure \ref{f:beta} and Table \ref{t:Mslopes}).  The value we find for the z=9-10 sample is somewhat flatter (Figure \ref{f:beta}), but this result is not especially significant and may change as more $z\sim9-10$ galaxies are identified and characterized. We have already noted that the derivation of stellar mass for current $z\sim9-10$ candidates is quite uncertain due to a number of issues (dust extinction, nebular lines of unknown strength).  However, we would expect the slopes we derive to be very similar however we deal with these uncertainties. The lack of a slope may be indicative that these galaxies are indeed the very first ones to be formed and are not yet completely virialized \citep[see Section 2 from][]{Stringer14}.ÊIn this case, the over-dense regions have collapsed --the lowest density ones first-- already forming the galactic system but not enough dynamical times have passed for the system to reach equilibrium and hence virial relations between mass and size. The different collapse times for different halo sizes also mean that the mass-size relation is not what is expected even for recently virialized systems. We reiterate, however, that the sample is small and the slope still uncertain. A larger $z\sim9-10$ sample would be needed to accurately determine this relation.

\begin{table}[htdp]
\caption{The slopes of the  mass-size relation. The lower-redshift values are from \protect\cite{van-der-Wel14} values,  derived for the late-type star-forming galaxies in their sample. The slopes for the \protect\cite{Mosleh12}, \protect\cite{Grazian12} and \protect\cite{Ono13} samples were derived by us based on the derived masses \protect\citep[using the ][conversions]{Stark13}, 
without assuming any evolution in the nebular emission EWs. The intercept ($R_0$) is fixed at a mass of $1. \times 10^{9} M_\odot$}
\begin{center}
\begin{tabular}{l l l l}
Redshift	& Intercept				& Slope 		& Reference\\
(z)		& ($R_0$, kpc)					& ($\beta$) 	& \\
\hline
\hline
0.25		& 2.72 $\pm$  0.04 			& 0.25 $\pm$ 0.02		& (1)\\
0.75		& 2.55 $\pm$  0.04			& 0.22 $\pm$ 0.01		& (1)\\ 
1.25		& 2.12 $\pm$  0.04			& 0.22 $\pm$ 0.01		&  (1)\\
1.75		& 1.82 $\pm$  0.04			& 0.23$\pm$ 0.01		&  (1)\\
2.25		& 1.50 $\pm$  0.04			& 0.22 $\pm$ 0.01		&  (1)\\
2.75		& 1.60 $\pm$  0.04			& 0.18 $\pm$ 0.02		&  (1)\\
\hline
6		& 0.75 $ \pm $ 0.18		& 0.14 $\pm$ 0.20 		& (2) \\
7		& 0.64 $ \pm $ 0.05			& 0.24 $\pm$ 0.08 		& (3)\\
7		& 0.27 $ \pm $ 0.07			& 1.35 $\pm$ 0.34 		& (4) \\
9-10		& 0.57 $ \pm $ 0.05 			& 0.12 $\pm$ 0.06 		& This work.\\
\hline
\end{tabular}
\end{center}
(1) \cite{van-der-Wel14}\\
(2) derived by us from the \cite{Mosleh12} data.\\ % please do not blame the nice people who did those original measurements.
(3) derived by us from the \cite{Grazian12} data.\\
(4) derived by us from the \cite{Ono13} data.\\
\label{t:Mslopes}
\end{table}%

\subsection{Redshift-Size Relationship}
\label{s:zr}

\begin{figure}
\begin{center}
\includegraphics[width=0.5\textwidth]{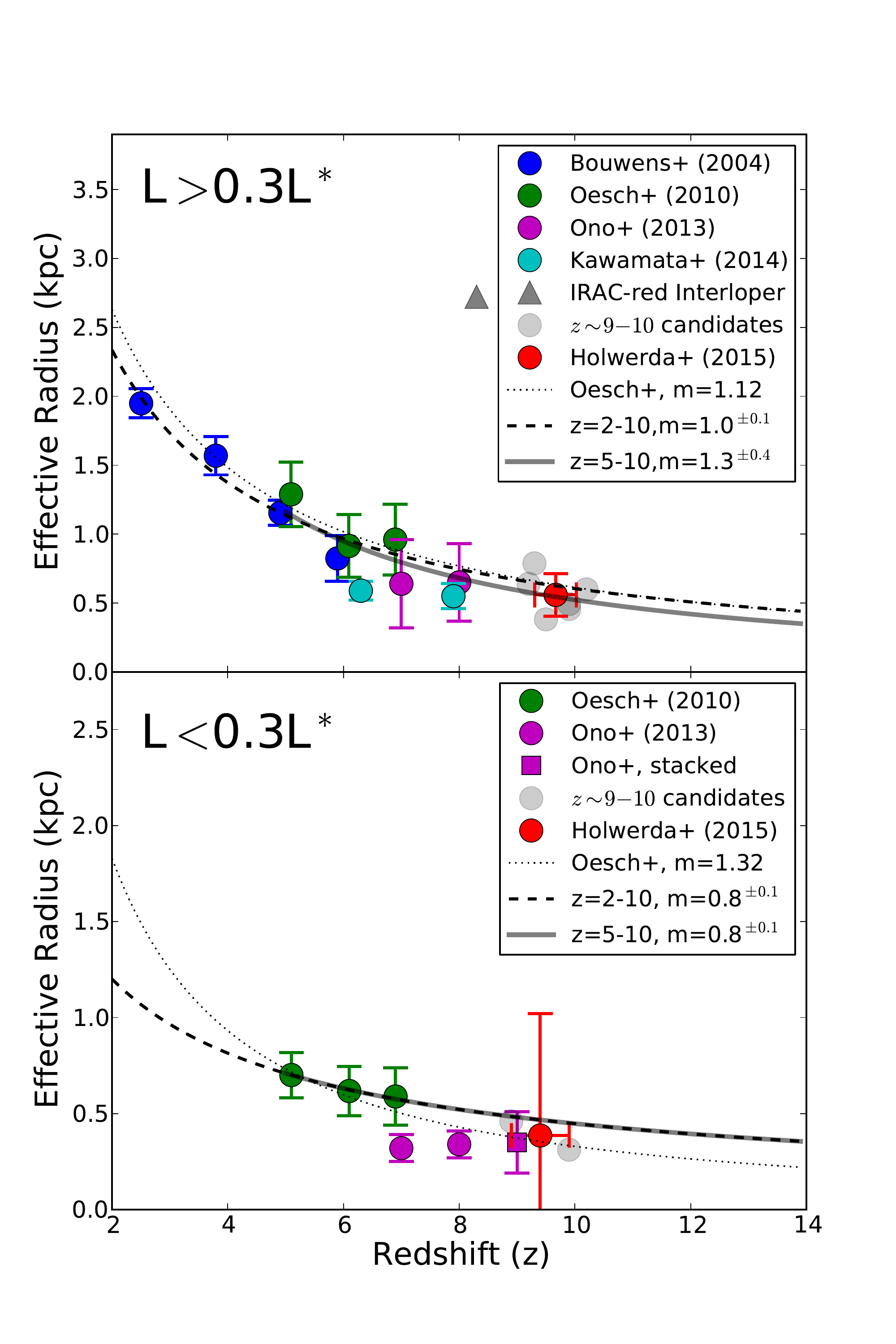}
\caption{The effective radius as a function of redshift for our sample for both 
bright ($L>0.3L^*_{z=3}$, top panel) and lower-luminosity galaxies ($L<0.3L^*_{z=3}$, bottom panel). 
For comparison, we show the mean sizes from earlier epochs from \protect\cite{Bouwens04}, \protect\cite{Oesch10b}, \protect\cite{Ono13}, and \protect\cite{Kawamata14}. 
The mean size of the six potential interlopers to a $z\sim9-10$ selection (see \S 4.1) is well above any expected relation at $z\sim9$. 
We do not include the \protect\cite{Bouwens11b} $z\sim2/z\sim12$ candidate as there is considerable doubt
as to whether it is at $z\sim12$  \protect\citep{Ellis13,Brammer13, Bouwens13,Capak13, Pirzkal13}.
The dotted line shows the best fits from \protect\cite{Oesch10b}. 
The dashed lines are our fits to the \cite{Bouwens04} and \cite{Oesch10b} values combined with our mean size constraints at $z\sim9-10$. We exclude the  \protect\cite{Ono13}, and \protect\cite{Kawamata14} points because these were derived using different methods.
The solid gray line the best fit for the high redshift ($z>5$) points alone.
The mean size of $L>0.3L^*_{z=3}$ galaxies scale as $(1+z)^{-1}$.}
\label{f:zReff}
\end{center}
\end{figure}

The discovery of a sample of luminous sources at $z\sim9-10$ provides us with additional leverage to constrain the size evolution of star-forming galaxies to $z\sim10$.
Figure \ref{f:zReff} shows the evolution of mean effective radius with redshift for luminous ($>0.3L^*_{z=3}$) and lower-luminosity ($<0.3L^*_{z=3}$) galaxies. 
It is important to be mindful of luminosity limits across redshift in examining size-redshift evolution \citep[see e.g.,][]{Cameron07}. For comparison, we include the mean size measurements from \cite{Bouwens04}, \cite{Oesch10b}, \cite{Ono13}, and \cite{Kawamata14}. We refer the reader to \cite{Shibuya15} for a discussion on the size evolution using parametrizations other than the mean (e.g., mode).

As the best-fit trend may be partially driven by the small uncertainties on the lower-redshift points, the value of our new $z\sim9-10$ size measurements for constraining the size evolution is somewhat limited assuming a fixed size-redshift scaling. Including our new $z\sim9-10$ size measurements and assuming a $(1+z)^{-m}$ scaling of size with redshift, the best-fit size-redshift scaling $m$ we find is 1.04 $\pm$ 0.09. Rederiving the scaling without our new constraints at $z\sim9-10$, we find 1.01 $\pm$ 0.10.
Previously, \cite{Bouwens04,Bouwens06} and \cite{Oesch10b} found a very similar dependence of mean size on redshift \citep[see also][]{Shibuya15}.
For lower-luminosity ($<0.3L^*_{z=3}$) galaxies, the evolution is much less certain ($m=0.8\pm0.1$), though the $(1+z)^{-1.32}$ relation from \cite{Oesch10b} also provides a reasonable fit.  Such a dependence is a generic expectation of theoretical models \citep[e.g.,][and others]{Somerville08,Wyithe11,Stringer14}. 

While we note only marginal improvements in our determination of the best-fit scaling including our new measurement, this is in the context of a model where galaxies are assumed to scale as a power of $1+z$ at all redshifts. It is conceivable that at early enough times galaxy sizes could scale differently (e.g., due to the impact of the UV ionizing background on gas cooling).  In this context, we have provided the first published constraints on the size evolution of luminous galaxies from $z\sim10$ to $z\sim8$.  

To illustrate, one can fit the evolution at the earliest epochs ($z\geq5$), where the statistical weight is no longer in the lowest redshift points. We do so with and without our $z\sim9-10$ constraint for both the luminous ($>0.3 L^*$) and lower-luminosity samples. We plot these fits to different redshift ranges in Figure \ref{f:zReff} and provide the best fit parameters in Table \ref{t:evolution}. Because so much weight is in the lower redshift points ($z<5$), the errors are  obviously the smallest if one includes the full redshift range ($z=2-10$). However, the inclusion of our latest high-redshift point improves the accuracy of the slope dramatically if one concerns oneself with the high-redshift evolution of sizes (Table \ref{t:evolution}).

\begin{table}[htdp]
\caption{The best-fit parameters, intercept and slope, for the luminous and lower-luminosity samples fit over different redshift ranges. If one includes the $z\sim9-10$ data in the high-redshift ($z>5$) fits, the accuracy improves significantly.}
\begin{center}
\begin{tabular}{l l l}
Redshift	& Intercept		& Slope \\
z		& $R_0 (z=4)$			& m	\\
\hline
\hline
$L>0.3L^*$	&				&	\\
2-8 & 1.38 $\pm$ 0.04 & 1.01 $\pm$ 0.10 \\ 
2-10 & 1.37 $\pm$ 0.04 & 1.04 $\pm$ 0.09 \\ 
   5-8 & 1.62 $\pm$ 0.60 & 1.64 $\pm$ 1.17 \\ 
   5-10 & 1.48 $\pm$ 0.26 & 1.32 $\pm$ 0.43 \\ 
\hline
$L<0.3L^*$	&				&	\\
   5-8 & 0.80 $\pm$ 0.03 & 0.71 $\pm$ 0.11 \\ 
   5-10 & 0.81 $\pm$ 0.03 & 0.76 $\pm$ 0.12 \\ 
\hline
\end{tabular}
\end{center}
\label{t:evolution}
\end{table}%

While the present study confirms that source size follows an approximate $(1+z)^{-1}$ scaling to very early times, it will be interesting to explore how the redshift-effective radius relation evolves for lower-mass galaxies as information on such systems become available in the future. 
For example, better relations between size and redshift, luminosity or mass will become available through expanded $z\sim9-10$ samples based on near-infrared photometric selections similar to the CANDELS ones using the future Frontier Fields program \citep[e.g.,][]{Kawamata14}, an extension to $z\sim9-10$ for the BoRG program \citep{Trenti14a} and in the very long term with the EUCLID \citep{EUCLID} or WFIRST \citep{WFIRST} satellites.

\section{Discussion}

In this paper we take advantage of six new bright $z\sim9-10$ candidate galaxies within CANDELS \citep{Oesch14} and their size information (1) to test their plausibility as $z\sim9-10$ sources and (2) to extend the study of the size-luminosity and size-mass relationship to $z\sim10$.
While most redshift $z\sim9-10$ candidate galaxies are unambiguously resolved  ($r_e > 0\farcs1$) with {\em HST} CANDELS or XDF {\em F160W} data (Figure \ref{f:stamps}), the brighter sources in our $z\sim9-10$ CANDELS sample are larger ($<r_e>=0\farcs13$) and better resolved than the fainter $z\sim10$ candidates in the HUDF/XDF ($<r_e>=0\farcs09$), allowing for a more optimal constraints on the sizes.

We find that the measured sizes can provide a useful test of the high-redshift nature of $z\sim9-10$ selections. In particular, we find excellent agreement between the sizes of our candidates and the extrapolation from lower redshift; interlopers to $z\sim9-10$ selections are in general $4\times$ larger (Figures \ref{f:conf} and \ref{f:zReff}). In the case of HST samples without IRAC coverage (e.g., the BORG[z9] HST/WFC3 pure-parallel survey), the size of the candidate high redshift galaxies can therefore potentially serve as an useful alternate constraint to select $z>9$ candidates. 

Secondly, we quantify the relationship between galaxy size and its luminosity at $z\sim9-10$.  The slope of the luminosity-size relation is lower than at $z=0$-6, but our sample is small and the uncertainties large.

% The star-formation remains the same
Thirdly, the absolute magnitude and effective radii of the $z=9$-10 galaxies imply a high average value of the star-formation surface density ($\rm \Sigma_{SFR}=4 ~ M_\odot~ yr^{-1}~ kpc^{-2}$, Figure \ref{f:SFR}), consistent with earlier estimates at z=4-8 \citep{Oesch10b,Ono13,Shibuya15}. 

% Mass-Size relation
Fourthly, we also explore the relationship between galaxy size and the stellar mass.  The mass-size relation slope (Figure \ref{f:beta}) for the $z\sim9-10$ sample is uncertain but flatter than the other comparison samples or the lower-redshift values reported in \cite{van-der-Wel14}.

Finally, for the first time, this resolved sample allows us to extend the redshift-size relation to $z\sim10$, confirming that $>0.3L^*_{z=3}$ galaxies follow an approximate $(1+z)^{-1}$ scaling as early as z=10. 

% JWST, ALMA and other targets
The mean sizes of these galaxies are informative for planning future extreme high-redshift observations with facilities such as EUCLID, WFIRST, JWST, ALMA and the various ELTs (see Figure \ref{f:model}), specifically their sizes and the implied star-formation surface densities.

\section*{Acknowledgements}

{
The authors are thankful to the referee for their meticulous feedback on the paper, which greatly helped in improving the manuscript.
}
% CANDELS
This work is based on observations taken by the CANDELS Multi-Cycle Treasury Program with the NASA/ESA HST, which is operated by the Association of Universities for Research in Astronomy, Inc., under NASA contract NAS5-26555 and HST GO-11563. 
% 3D HST
This work is based on observations taken by the 3D-HST Treasury Program (GO 12177 and 12328) with the NASA/ESA HST, which is operated by the Association of Universities for Research in Astronomy, Inc., under NASA contract NAS5-26555.
This research has made use of NASA's Astrophysics Data System.
This research made use of Astropy, a community-developed core Python package for Astronomy \citep{Astropy-Collaboration13a}. This research made use of matplotlib, a Python library for publication quality graphics \citep{Hunter07}. PyRAF is a product of the Space Telescope Science Institute, which is operated by AURA for NASA. This research made use of SciPy \citep{scipy}. 
We acknowledge the support of ERC grant HIGHZ \#227749, and a NWO ``Vrije Competitie" grant 600.065.140.11N211 and the NL-NWO Spinoza.

%\bibliographystyle{apj} 
%\bibliography{/Users/bholwerd/Desktop/Science/Bib/Bibliography}

% \end{document}
\appendix

\section{Size Evolution Model}

\cite{Wyithe11} present a simple model based on the luminosity function slope ($\alpha=1-a$) and the size evolution ($m$) to estimate the sizes of galaxies for future observations and observatories. They arrive at a model for the galaxy size which depends on luminosity ($m_{AB}$), redshift ($z$) as follows (their equation 9):

\begin{equation}
\label{eq:model}
r_e = R_0 \left( {D_L(z) \over D_L (z_0)} \right)^{2 \over 3(1+a)} 10^{m_{AB,0} - m_{AB} \over 7.5(a+1)} \left( {1+z \over 1+z_0} \right)^{-m},
\end{equation}
\noindent where $R_0$, and $m_{AB,0}$ are normalization parameters determined at a later epoch ($z_0$). They adopt the mean of some of the \cite{Oesch09} results. We adopt the $z_0=8$ values:  $R_0 = 0.4$, $m_{AB,0} = 28.1$, $\alpha=-2.27$ \citep{Bouwens14} and $m=-1$ in Figure \ref{f:model}.

For the observatories, they assume diffraction limited observations, i.e.:
\begin{equation}
\label{eq:telescope}
\Delta \theta = {  1.22\lambda \over D_{tel}} \simeq 0.085 \left({1+z\over7}\right) \left({D_{tel}\over 2.5}\right)^{-1},
\end{equation}
\noindent for the wavelength of Lyman-$\alpha$ and a fiducial $m_{AB}= 28$ source. 

Figure \ref{f:model} shows our comparison to these first expectations to our $z\sim9-10$ objects. It shows we do slightly better than a simple diffraction estimate because the CANDELS data is drizzled and we have a good PSF model in hand. The model's prediction based on the values above match those of our size measurements for the given AB luminosities. 

\begin{figure}
\begin{center}
\includegraphics[width=0.5\textwidth]{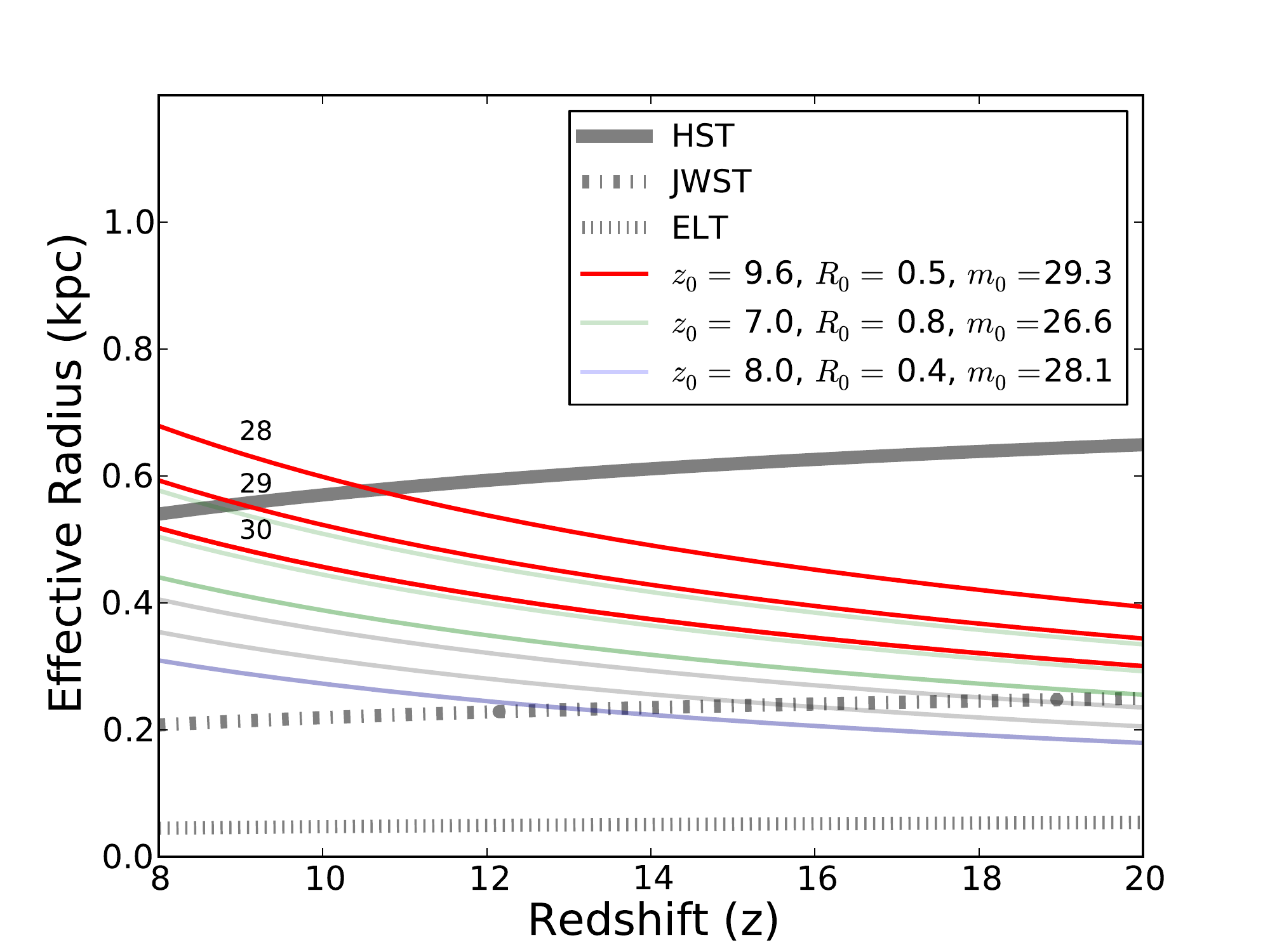}
\caption{The expected size-redshift relation from the simple model of \protect\cite{Wyithe11}, normalized with different size distribution observed at high redshift ($R_0$ for a $m_0$ galaxy at $z_0$). The blue and green lines are z=7 and z=8 normalizations based on \protect\cite{Oesch09} for $m_{0} = 28, 29$, and $30$ respectively. The red lines are based on our $z\sim9-10$ sample presented here. How well one can expect to resolve the earliest epoch galaxies with future not only depends on the adopted parametrization of size-evolution but the high-redshift normalization. }
\label{f:model2}
\end{center}
\end{figure}

Starting from the model presented in  \cite{Wyithe11}, we can extrapolate the $z\sim9-10$ sample (similar to their Figure 2). 
Figure \ref{f:model2} shows the model extrapolation from our $z\sim9-10$ objects. The  simple model from  \cite{Wyithe11}, suffices to predict galaxy sizes at earliest times, it needs to be anchored to the highest-redshift measurements available if one is to successfully plan observations with future observatories such as JWST and ELT (also shown in Figure \ref{f:model}). 

Based on the sizes presented in \cite{Oesch09} ($z_0=7$, green lines, $z_0=8$ blue lines), one would not have expected HST to resolve the fainter of our sources ($m_{AB}\sim28$). However, with these $z\sim9-10$ sources confirmed, it appears possible that HST may still discover and resolve some rare $z=10-11$ $m_{AB}=29$ objects. Conversely, JWST and ELT planning will have to take more extended galaxies into account (e.g., NIRspec slit width etc).

%@archiver{holwerda_f1.pdf, holwerda_f2.pdf, holwerda_f6.pdf}

\end{document}